\documentclass[a4paper,11pt]{article}
\pdfoutput=1 
\usepackage{jcappub} 

\makeatletter
\gdef\@fpheader{}
\g@addto@macro\bfseries{\boldmath}
\makeatother

\usepackage[T1]{fontenc} 
\usepackage{lmodern} 

\usepackage{nowidow} 
\usepackage{float} 

\hypersetup{pdftitle={Primordial black holes from a curvaton scenario with strongly non-Gaussian perturbations},pdfauthor={3 authors}}

\def\baq{\begin{eqnarray}}
\def\eaq{\end{eqnarray}}

\def\beq{\begin{equation}}
\def\eeq{\end{equation}}

\newcommand{\diffd}{\mathrm{d}} 

\title{Primordial black holes from a curvaton scenario with strongly non-Gaussian perturbations}

\author[a]{Andrew Gow,}
\author[b,c]{Tays Miranda,}
\author[c,b]{Sami Nurmi\;}

\affiliation[a]{Institute of Cosmology \& Gravitation, University of Portsmouth, Dennis Sciama Building, Burnaby Road, Portsmouth, PO1 3FX, United Kingdom}
\affiliation[b]{Helsinki Institute of Physics, P.O. Box 64, FIN-00014 University of Helsinki, Finland}
\affiliation[c]{Department of Physics, P.O.Box 35 (YFL), FIN-40014 University of Jyväskylä, Finland}

\emailAdd{andrew.gow@port.ac.uk}
\emailAdd{tays.miranda@helsinki.fi}
\emailAdd{sami.t.nurmi@jyu.fi}

\abstract{We investigate the production of primordial black holes (PBHs) in a mixed inflaton--curvaton scenario with a quadratic curvaton potential, assuming the curvaton is in de Sitter equilibrium during inflation with $\langle \chi\rangle =0$. In this setup, the curvature perturbation sourced by the curvaton is strongly non-Gaussian, containing no leading Gaussian term.  We show that for $m^2/H^2\gtrsim 0.3$, the curvaton contribution to the spectrum of primordial perturbations on CMB scales can be kept negligible but on small scales the curvaton can source PBHs. In particular, PBHs in the asteroid mass range $10^{-16}M_{\odot}\lesssim M\lesssim 10^{-10}M_{\odot}$ with an abundance reaching $f_{\rm PBH} = 1$ can be produced when the inflationary Hubble scale $H\gtrsim 10^{12}$ GeV and the curvaton decay occurs in the window from slightly before the electroweak transition to around the QCD transition.}

\begin{document}
\maketitle
\flushbottom

\section{Introduction}
\label{sec:section1}

The study of primordial black holes (PBHs) formed via gravitational collapse of large primordial density fluctuations was initiated over $50$ years ago~\cite{Zeldovich:1967lct,Hawking:1971ei}.  Already in~\cite{Chapline:1975ojl} it was proposed that PBHs could form a cold dark matter component in the universe. The possibility that PBHs of mass $M\gtrsim 10^{-10}M_{\odot}$ would constitute all of the dark matter is already ruled out by constraints from lensing, dynamical effects, structure formation and gravitational waves~\cite{Carr:2020gox}. In the asteroid mass window, $10^{-16}M_{\odot}\lesssim M\lesssim 10^{-10}M_{\odot}$ the constraints are uncertain and it is possible that PBHs with masses in this range could constitute a dominant dark matter component~\cite{Carr:2020gox}. Even a subdominant PBH contribution to dark matter can have characteristic observational imprints, such as gravitational waves from PBH mergers testable with LIGO--Virgo--KAGRA data~\cite{Bird:2016dcv,Clesse:2016vqa,Sasaki:2016jop,Gow:2019pok,Franciolini:2021tla}. Observational signals associated to PBHs are among the very few ways to probe small scale primordial perturbations and the process responsible for their generation. 

Several mechanisms for producing PBHs have been investigated in the literature, including perturbations produced during inflation~\cite{Ivanov:1994pa,Garcia-Bellido:1996mdl,Ivanov:1997ia,Leach:2000ea,Drees:2011hb,Drees:2011yz,Garcia-Bellido:2017mdw,Domcke:2017fix,Kannike:2017bxn,Germani:2017bcs,Motohashi:2017kbs,Ballesteros:2017fsr,Hertzberg:2017dkh,Pi:2017gih,Kohri:2018qtx,Biagetti:2018pjj,Dalianis:2018frf,Ballesteros:2018wlw,Georg:2019jld,Pi:2019ihn,Germani:2018jgr,Carr:2018poi,Kamenshchik:2018sig}, cosmic strings ~\cite{Hawking:1987bn,Garriga:1993gj,Caldwell:1995fu}, phase transitions~\cite{Jedamzik:1996mr,Byrnes:2018clq,Chakraborty:2022mwu}, dark matter clumps~\cite{Shandera:2018xkn}, and bouncing cosmologies~\cite{Chen:2016kjx,Quintin:2016qro}. In this work, we focus on the curvaton scenario~\cite{Lyth:2001nq, Moroi:2001ct, Enqvist:2001zp, Linde:1996gt, Mollerach:1989hu}. Several authors have already investigated PBH formation in curvaton models ~\cite{Klimai:2012sf,Young:2013oia,Kawasaki:2012wr,Pi:2021dft,Meng:2022low} and other spectator scenarios~\cite{Carr:2016drx,Garcia-Bellido:2016dkw,Carr:2017edp,Ezquiaga:2017fvi,Espinosa:2017sgp,Cable:2023lca} where the curvature perturbation can acquire non-Gaussian contributions. The effect of this non-Gaussianity is typically treated perturbatively, using the parameters $f_\text{NL}$ \textit{etc}.~\cite{Bullock:1996_Non-Gaussian,Young:2015_Influence,Yoo:2019_Abundance,Taoso:2021_Non-gaussianities,Young:2022_non-G}. Recently, a general non-perturbative formalism for investigating the PBH abundance in the presence of non-Gaussian perturbations has been developed \textit{e.g.}~in~\cite{Gow:2022jfb,Ferrante:2022mui,Ferrante:2023bgz}. To our knowledge, the previous analyses of PBH production in the curvaton scenario have however focused on the case where the curvaton field has a non-vanishing mean value during inflation, $|\langle \chi \rangle| \gg H$, and the curvature perturbation contains a leading Gaussian part proportional to $\delta \chi/\langle \chi \rangle \ll 1$. The non-Gaussianities can then be accommodated using a truncated  expansion around the Gaussian part, although see~\cite{Young:2013oia} for a discussion of limitations of this method when $\langle \chi \rangle \gg H$. 

In this work, we will study PBH formation in the curvaton setup with a quadratic potential,  assuming $\langle \chi \rangle = 0$. This is the equilibrium configuration of a light spectator during de Sitter inflation~\cite{Starobinsky:1994bd}, provided the potential is minimized at $\chi=0$. Even if one starts from an initial configuration with a non-zero $\langle \chi \rangle$, the distribution is rapidly driven towards the equilibrium during de Sitter inflation~\cite{Starobinsky:1994bd,Enqvist:2012xn}\footnote{When the inflationary solution deviates from de Sitter, the relaxation towards the de Sitter equilibrium may happen at a slower rate, or there may also be parameter combinations for which the equilibrium is not reached~\cite{Hardwick:2017fjo}.}.  For $\langle \chi \rangle =0$, perturbations of the curvaton energy density obey the distribution of a squared Gaussian field and are large, $\langle\delta \rho_{\chi}^2\rangle/\langle \rho_{\chi}\rangle^2 = 2$~\cite{Lyth:2001nq}. Consequently, the curvature perturbation component sourced by the curvaton has no leading Gaussian part and it can not be expanded in small perturbations. Therefore, one needs to consider the full non-linear and non-Gaussian solution for the curvature perturbation when investigating the PBH formation. The curvature perturbation $\zeta$ on CMB anisotropy scales is Gaussian to a high precision~\cite{Planck:2019kim} and contributions sourced by $\delta\rho_{\chi}/\langle\rho_{\chi}\rangle$ must be strongly suppressed on these scales. If the curvaton spectrum is sufficiently blue-tilted, it can however give a dominant contribution to $\zeta$ on small scales $k\gg {\rm Mpc}^{-1}$ where there are no constraints on non-Gaussianity. We focus on such a setup, investigating a mixed inflaton--curvaton scenario with a strongly blue tilted curvaton component which dominates the curvature perturbation on small scales and sources a strongly non-Gaussian $\zeta$ with no leading Gaussian component. On the CMB scales, we require that the spectrum of $\zeta$ is dominated by a Gaussian inflaton component and contributions from the curvaton are suppressed to the $10^{-11}$ level at the CMB pivot scale $k_{*} =0.05\ {\rm Mpc}^{-1}$. We note that PBHs in a partially related phenomenological setup with a Gaussian blue tilted spectator curvature perturbation component was investigated in~\cite{Carr:2017edp}.

We use the $\delta N$ approach~\cite{Starobinsky:1985ibc,Salopek:1990jq,Sasaki:1995aw,Wands:2000dp} to obtain a non-linear solution for the superhorizon scale curvature perturbation. We expand the curvaton field $\chi({\bf x})$ in spherical harmonics and, following~\cite{Gow:2022jfb}, truncate the expansion to the monopole when considering fluctuations relevant for PBHs. Within this approximation, we compute the probability distribution for $C_{l}(r)$ which determines the compaction function $C(C_{l}(r))$. The compaction function $C(r)$ equals the comoving gauge density contrast smoothed over a radius $r$. Using $C_{\rm c} =0.55$~\cite{Musco:2008hv} as the PBH collapse threshold and modeling the PBH mass with the collapse parameters obtained in~\cite{Musco:2020jjb} for the power law spectrum, we explore the fraction of dark matter in PBHs $f_{\rm PBH}$ and the mass distribution $f(M)$ as functions of the curvaton model parameters. We show that the curvaton scenario with the quadratic potential and the equilibrium configuration $\langle \chi \rangle = 0$ can lead to very efficient PBH production. In particular, we find that the scenario can produce asteroid mass PBHs, $10^{-16}M_{\odot}\lesssim M\lesssim 10^{-10}M_{\odot}$, with $f_{\rm PBH} = 1$ when the inflationary Hubble scale $H\gtrsim 10^{12}$ GeV, the curvaton mass $m^2/H^2 \gtrsim 0.3$ and the curvaton decay occurs in the window from slightly before the electroweak transition to around the QCD transition.  

The paper is organised as follows. In Sec.~\ref{sec:inflation_setup} we present the setup and in Sec.~\ref{sec:PDF} we compute the probability distribution for the compaction function. In Sec.~\ref{sec:PBH} we collect the expressions for the PBH abundance $f_{\rm PBH}$ and the mass distribution $f(M)$. In Sec.~\ref{sec:results} we present our main results and summarise the discussion in Sec.~\ref{sec:summary}. 

\section{The mixed inflaton--curvaton setup}
\label{sec:inflation_setup}

We investigate the PBH abundance in a mixed inflaton--curvaton scenario where the inflaton generates the Gaussian, nearly scale invariant curvature perturbations on CMB scales and curvaton-sourced perturbations dominate on small scales relevant for PBH formation. We assume the quadratic curvaton potential 
\beq
\label{Vchi}
V(\chi)=\frac{1}{2}m^2 \chi^2~.
\eeq
We further assume that the curvaton distribution during inflation follows the de Sitter equilibrium result with a vanishing mean value $\langle \chi \rangle = 0$~\cite{Starobinsky:1994bd}. Ebadi \textit{et al.}~\cite{Ebadi:2023xhq} recently examined the production of gravitational waves for this case. For the quadratic potential, the curvaton $\chi$ is a Gaussian field and for $m^2/H^2 < 3/2$ its spectrum at the end of inflation on superhorizon scales is given by the standard power-law expression~\cite{Bunch:1978yq}
\beq
\label{eq:Pchipowerlaw}
{\cal P}_{\chi}(k)= \left(\frac{H}{2\pi}\right)^2 \left(\frac{k}{k_{\rm end}}\right)^{3-2\nu}\frac{2^{2\nu-1}\Gamma(\nu)^2}{\pi}~,\qquad\nu=\sqrt{\frac{9}{4}-\frac{m^2}{H^2}}~.
\eeq 
We denote the Hubble scale at the end of inflation by $H\equiv H(t_{\rm end})$, and $k_{\rm end} = a(t_{\rm end}) H(t_{\rm end})$ is the mode exiting the horizon at the end of inflation. Perturbations of the curvaton energy density
\beq
\label{rhochi}
\frac{\delta \rho_{\chi}({\bf x})}{\langle \rho_{\chi}\rangle} = \frac{\rho_{\chi({\bf x})}-\langle \rho_{\chi}\rangle}{\langle \rho_{\chi}\rangle}=\frac{\chi^2({\bf x})}{\langle\chi^2\rangle}-1, 
\eeq
obey the statistics of a Gaussian squared quantity, and the perturbations are large since $\langle\delta \rho_{\chi}^2({\bf x})\rangle/\langle \rho_{\chi}\rangle^2 = 2$. Consequently, any contribution to the curvature perturbation sourced by $\delta \rho_{\chi}/\langle \rho_{\chi}\rangle$ must be suppressed on the large scales $k\lesssim {\rm Mpc^{-1}}$ probed by the CMB and LSS data~\cite{Planck:2019kim}. 

Using the $\delta N$ formalism~\cite{Starobinsky:1985ibc,Salopek:1990jq,Sasaki:1995aw,Wands:2000dp}, the superhorizon curvature perturbation $\zeta$ in the mixed scenario with inflaton sourced perturbations in the radiation component and the perturbed curvaton component obeys the non-linear equation 
\cite{Sasaki:2006kq}
\begin{align}\label{eq:mastercurv}
e^{4\zeta}-\Omega_{\chi}e^{3\zeta_\chi}e^{\zeta}+\left(\Omega_{\chi}-1\right)e^{4\zeta_{\rm r}}=0~. 
\end{align}
The individual curvature perturbations of the radiation $\zeta_{\rm r}$ and curvaton $\zeta_{\chi}$ fluids, and $\Omega_{\chi}$ are given in terms of spatially flat gauge quantities by the expressions,    
\beq
\zeta_{\rm r}({\bf x}) = \frac{1}{4}{\rm ln}\frac{\rho_{\rm r}({\bf x})}{\langle \rho_{\rm r} \rangle}~,\qquad
\zeta_{\rm \chi}({\bf x}) = \frac{1}{3}{\rm ln}\frac{\rho_{\chi}({\bf x})}{\langle \rho_{\chi} \rangle}~,\qquad
\Omega_{\chi} = \frac{\langle \rho_{\chi}\rangle }{\langle\rho_{\rm r}\rangle+\langle\rho_{\chi}\rangle}~.
\eeq
Both $\zeta_{\rm r}$ and $\zeta_{\chi}$ are separately conserved,  \textit{i.e.}~constant in time.  For the quadratic potential Eq.~\eqref{Vchi},  the curvaton component $\zeta_{\chi}$ can be written in terms of the field $\chi$ as 
\beq
\label{eq:zetachi}
\zeta_{\chi}=\frac{1}{3}{\rm ln}\frac{\chi^2}{\langle \chi^2\rangle} ~.
\eeq
Here and in the rest of the text we use $\chi \equiv \chi_{\rm end} $ to denote the curvaton field  at the end of inflation.  

The solution for the fourth order algebraic equation (\ref{eq:mastercurv}) can be written as~\cite{Sasaki:2006kq} 
\baq
\label{eq:zetasolfull}
\zeta &=& \zeta_{\rm r} + {\rm ln}\left(K^{1/2}\left(\frac{4-\Omega_{\chi}}{12}\right)^{1/3}\left[1+\left(\frac{3 \Omega_{\chi}K^{-3/2}e^{3(\zeta_{\chi}-\zeta_{\rm r})}}{4-\Omega_{\chi}}-1\right)^{1/2}\right] \right)~,\\\nonumber
K&=&\frac{1}{2}\left[P^{1/3}+\frac{4\Omega_{\chi}-4}{4-\Omega_{\chi}}\left(\frac{12}{4-\Omega_{\chi}}\right)^{1/3}P^{-1/3} \right]~,\\\nonumber
P&=&\left( 
\frac{3 \Omega_{\chi}e^{3(\zeta_{\chi}-\zeta_{\rm r})}}{4-\Omega_{\chi}}
\right)^2+\left[\left( 
\frac{3 \Omega_{\chi}e^{3(\zeta_{\chi}-\zeta_{\rm r})}}{4-\Omega_{\chi}}
\right)^4+\frac{12}{4-\Omega_{\chi}}\left(\frac{4-4\Omega_{\chi}}{4-\Omega_{\chi}}\right)^3\right]^{1/2}~.
\eaq
Here the only time dependent quantity is the curvaton density parameter $\Omega_{\chi}$.  Assuming the curvaton is subdominant at the onset of oscillations, which we define as $H(t_{\rm osc}) \equiv m$, the density parameter for $t\gg t_{\rm osc}$ can be written as 
\beq
\label{eq:Omega}
\Omega_{\chi}=\frac{\Omega_{\chi,{\rm osc}}}{\Omega_{\chi,{\rm osc}}+\frac{a_{\rm osc}}{a}}\approx\frac{0.136\langle\chi^2\rangle}{0.136\langle\chi^2\rangle +\frac{a_{\rm osc}}{a}M_{\rm P}^2}~.
\eeq
Here we used that $\rho_{\chi,{\rm osc}}\approx (1/2)m^2 0.816 \chi^2$ which follows from solving the curvaton equation of motion in a radiation dominated universe.  We approximate the curvaton decay as an instant process at $H(t_{\rm dec}) = \Gamma$.  Within this approximation $\zeta \equiv \zeta(t > t_{\rm dec}) = \zeta(t_{\rm dec})$ is obtained by evaluating Eq.~\eqref{eq:zetasolfull} at the moment $t_{\rm dec}$.

We assume the perturbation of the radiation component $\zeta_{\rm r}$ is entirely sourced by the inflaton and uncorrelated with the curvaton, $\langle \zeta_{\rm r} \zeta_{\chi} \rangle = 0$. We further assume $\zeta_{\rm r}$ obeys Gaussian statistics with the power law spectrum 
\beq
\label{eq:Pzetar}
 {\cal P}_{\zeta_{\rm r}}(k)=A_{\rm r}\left(\frac{k}{k_*}\right)^{n_{\rm r}-1}~,
\eeq
where $k_* =0.05\ {\rm Mpc}^{-1}$. As explained above, we want to realise a scenario where the Gaussian inflaton sourced perturbations $\zeta_{\rm r}$ dominate the two point function of $\zeta$ on large scales and generate the observed CMB spectrum with $P_{\zeta}(k_*)=A_{\rm s}=2.10 \times 10^{-9}, n_{\rm s}= 0.965$~\cite{Planck:2018vyg}. Correspondingly, contributions from $\zeta_{\chi}$ to the spectrum of $\zeta$ must be sufficiently suppressed. To this end, we require 
\beq
\label{eq:Pzetachicond}
 {\cal P}_{\zeta_{\chi}}(k_*) < 2 \times 10^{-11}~, 
\eeq
which, as we show in Appendix~\ref{app:spectrum_CMB}, allows us to obtain the observed CMB spectrum with $A_{\rm r}\sim 10^{-9}$ and $|n_{\rm r}-1|\sim 0.01$, assuming $\Omega_{\chi}$ $\lesssim 0.9$. Note that the underlying computation is somewhat non-trivial as $\zeta$ is a strongly non-linear function of the non-Gaussian $\zeta_{\chi}$. The condition (\ref{eq:Pzetachicond}) constrains the curvaton mass from below as we show in detail in Appendix~\ref{app:spectrum_zetachi}, and implies a strongly blue tilted spectrum both for the curvaton field Eq.~\eqref{eq:Pchipowerlaw} and for $\zeta_{\chi}$. On small scales, $k\gg {\rm Mpc}^{-1}$, the connected correlators of $\zeta$ are dominated by $\zeta_{\chi}$ which, as we show below, can lead to efficient formation of PBHs. For the maximal inflationary Hubble scale $H \approx 5.2 \times 10^{13}\;{\rm GeV}$ consistent with the observational bound on the tensor-to-scalar ratio $r_{\rm T} < 0.044$~\cite{Tristram:2020wbi}, Eq.~\eqref{eq:Pzetachicond} implies $m^2/H^2 \gtrsim 0.29$, assuming instant transition from inflation to radiation domination. The lower bound on $m^2/H^2$ slowly grows as a function of decreasing $H$. For example, for $H = 1.6\times 10^{11}\ {\rm GeV}$, Eq.~\eqref{eq:Pzetachicond} implies $m^2/H^2 \gtrsim 0.31$, see Fig.~\ref{fig:m2overh2plot} in Appendix~\ref{app:spectrum_CMB}.

\section{Probability distribution of density fluctuations}
\label{sec:PDF}

The central quantity in determining if an overdense region collapses into a primordial black hole is the compaction function $C(r)$~\cite{Shibata:1999zs, Harada:2015yda,Yoo:2018kvb} which for spherical overdensities equals the comoving density contrast coarse-grained with a top hat window function over a spherical volume of comoving radius $r$. Denoting the background equation of state by $w =\langle p\rangle /\langle \rho\rangle$,  the expression for $C(r)$ can be written as~\cite{Musco:2018rwt,Young:2019yug}  
\beq
\label{eq:C(r)}
C(r)=C_l(r) -  \frac{1}{2f(w)}C_l(r) ^2~,
\eeq
where
\beq
\label{eq:fw}
f(w) = \frac{6(1+w)}{5+3w}~,
\eeq
and $C_l(r)$ is determined by the curvature perturbation $\zeta$ as 
\beq
\label{eq:Cl(r)def}
C_l(r) = - f(w) r\zeta'(r)~.
\eeq
Here the prime denotes a derivative with respect to $r$.

Around the high density peaks relevant for the PBH formation, spherical symmetry can be expected to be a reasonable first approximation.  We implement the approximation following~\cite{Gow:2022jfb} by expanding the Gaussian field $\chi$ in spherical harmonics 
\beq
\label{eq:chifull}
\chi(\textbf{x}) =\int \frac{\diffd\textbf{k}}{(2\pi)^3}{\chi}_{{\bf k}}4\pi \sum_{l,m}i^{l}j_{l}(kx)Y_{lm}({\bf\hat{x}})Y_{lm}^{\star}({\bf\hat{k}})~,
\eeq
and retaining only the leading monopole term of the expansion
\beq
\label{eq:chir}
\chi(r) =\int \frac{{\diffd}{\bf k}}{(2\pi)^3}j_0(kr)\chi_{\bf k}~.  
\eeq
Here  $j_{0}(z)=\sin(z)/z$.  Since Eq.~\eqref{eq:chir} is a linear map from $\chi(\bf{x})$,  the field $\chi(r)$ and its derivative $\chi'(r)$ are Gaussian fields with the joint probability distribution given by 
\beq
\label{eq:Pchichiprime}
P_{\chi\chi'}(\chi,\chi')=\frac{1}{2\pi \sqrt{|\Sigma|}} {\rm exp}\left(-\frac{1}{2}X^{T}\Sigma^{-1}X\right)~,~~X^{T} = (\chi,\chi')~,~~\Sigma^{-1}=\begin{pmatrix}
  \sigma^{2}_{\chi\chi} & \sigma^{2}_{\chi\chi^{\prime}} \\[0.3em]
  \sigma^{2}_{\chi\chi{\prime}} & \sigma^{2}_{\chi{\prime}\chi{\prime}} \\
\end{pmatrix}~.
\eeq
The components of the covariance matrix depend on $r$ and can be written as 
\baq
\label{eq:Sigmacomp}
\sigma^2_{\chi\chi}(r)&=&\langle\chi(r)\chi(r)\rangle=\int \diffd{\rm ln}k\; j_{0}^{2}(kr){\cal P}(k) \;,\\
\sigma^2_{\chi\chi'}(r)&=&\langle\chi(r)\chi'(r)\rangle=\int \diffd{\rm ln}k\; j_{0}'(kr)j_{0}(kr){\cal P}(k) \;,\\
\sigma^2_{\chi'\chi'}(r)&=&\langle\chi'(r)\chi'(r)\rangle = \int \diffd{\rm ln}k\; \left(j_{0}'(kr)\right)^{2}{\cal P}(k)~, 
\eaq
where a prime denotes a derivative with respect to $r$ and ${\cal P}(k)$ is the power spectrum of the full curvaton field $\chi({\bf x})$
\beq
\label{eq:Pdef}
\langle \chi_{\bf k} \chi_{\bf k'}\rangle = (2\pi)^3 \delta({\bf k}+{\bf k}')\frac{2\pi^2}{k^3}{\cal P}(k)~. 
\eeq
In our case ${\cal P}(k)$ is given by Eq.~\eqref{eq:Pchipowerlaw}.
We denote the variance of the full field $\chi({\bf x})$ by $\sigma^2$,
\beq
\label{eq:s2full}
\sigma^2\equiv \langle \chi^2 ({\bf x})\rangle = \int \diffd{\rm ln}k\;{\cal P}(k)~. 
\eeq

We proceed to substitute $\chi({\bf x}) \rightarrow \chi(r)$ in Eqs.~\eqref{eq:zetachi} and \eqref{eq:zetasolfull} in places where the field $\chi({\bf x})$ appears with no brackets,  but  evaluate the background quantities that depend on $\langle \chi^2 \rangle$ in Eq.~\eqref{eq:zetasolfull} using the  variance of the full field (\ref{eq:s2full}).  Using $\langle \chi(r)^2\rangle < \langle \chi({\bf x})^2 \rangle$ would give a higher probability for larger $\zeta_{\chi}$ values and hence enhance the PBH abundance but it is hard to quantify to  what extent this is a spurious effect of the monopole truncation.  We will therefore conservatively use $\langle \chi({\bf x})^2 \rangle$ in the background quantities.  

In this setup, Eq.~\eqref{eq:Cl(r)def} takes the form  
\beq
\label{eq:Cl(r)}
C_l(r) = - r f(w)\chi'(r)\partial_{\chi}\zeta(\chi(r))~, 
\eeq
where $\partial_{\chi} \zeta \equiv \frac{\partial \zeta}{\partial\chi} $ is obtained by differentiating Eq.~\eqref{eq:zetasolfull}.  The probability distribution of $C_{l}$ is given by 
\beq
P_{C_l}(C_l,r) = \int \int \diffd\chi  {\rm d}\chi'\ P_{\chi\chi'}(\chi,\chi')\delta\Big[C_l+
r f(w)\chi'(r)
\partial_{\chi}\zeta(\chi(r))\Big]~.
\eeq
Carrying out the integral over $\chi'$,  we obtain 
\baq
\label{eq:fullpdf}
P_{C_l}(C_l,r) &=&\frac{1}{2 \pi f(w) r |\Sigma(r)|^{1/2}}\times\\\nonumber
&&\int \frac{\diffd\chi}{|\partial_{\chi}\zeta|}{\rm exp}\left[-\frac{1}{2|\Sigma(r)|}\left(\sigma^2_{\chi'\chi'}(r)\chi^2+\frac{2\sigma^2_{\chi\chi'}(r)\chi C_l}{f(w)r\partial_{\chi}\zeta}+\frac{\sigma^{2}_{\chi\chi}(r)C_l^2}{(f(w)r\partial_{\chi}\zeta)^2}\right)\right]~,
\eaq
where the remaining integral needs to be computed numerically. 

\section{Expression for the PBH abundance}
\label{sec:PBH}

The mass of a PBH formed by a collapsing region with compaction $C$ can be approximated by  
\beq
\label{eq:Mscaling}
M =K  M_{\rm H}\left(C-C_{\rm c}\right)^{\gamma}~,
\eeq
obtained by fitting to numerical simulations~\cite{Choptuik:1992jv,Evans:1994pj,Niemeyer:1997mt,Niemeyer:1999ak,Musco:2004ak,Musco:2012au,Musco:2008hv}. Here  $M_{\rm H} = 4\pi/3 H^{-3} \rho$ is the mass within a Hubble volume at the collapse time, $\gamma$ depends on the equation of state, and $K$ and the collapse threshold $C_{\rm c}$ depend on both the equation of state and the shape of the collapsing overdensity. In a radiation dominated universe $\gamma \approx 0.36$. For monochromatic PBHs formed during radiation domination from a Gaussian $\zeta$ with the spectrum ${\cal P}_{\zeta}(k)\propto \delta (k-k_*)$, $K={\cal O}(1)$ and $C_{\rm c} \approx 0.59$, and the overdensity peaks at the comoving scale $r_{*} \approx 2.74/k_* $~\cite{Musco:2018rwt,Young:2019osy,Germani:2018jgr,Escriva:2020tak,Musco:2020jjb}. For a Gaussian $\zeta$ with a nearly scale invariant power-law spectrum and radiation domination, $ K\approx 4$ and $C_{\rm c} \approx 0.55$, and the overdensity generated by a mode $k_*$ peaks at  $r_{*} = 4.49/k_* $~\cite{Musco:2008hv,Musco:2020jjb}.

There are no existing numerical collapse simulations corresponding to our case with a strongly non-Gaussian $\zeta$ which does not have a leading Gaussian part. We adopt a phenomenological approach and set the collapse parameters equal to the Gaussian power-law results in radiation domination, $\gamma = 0.36$, $C_{\rm c} = 0.55$, and $K=4$. Therefore, we will compute the PBH mass using   
\beq
\label{eq:Mpbh}
M(C,r) = 4 M_{\rm H}(r)\left(C -0.55\right)^{0.36}~.
\eeq
We evaluate $M_{\rm H}(r)$ at the horizon entry of the smoothing scale $a(t_{\rm r}) H(t_{\rm r}) r = 1$, 
\beq
\label{eq:MH}
M_{\rm H}(r) = \left.\frac{4 \pi}{3} H(t_{\rm r})^{-3}\rho(t_{\rm r})\right|_{a(t_{\rm r}) H(t_{\rm r}) r =1}~.
\eeq
In our setup, the curvaton contribution to the energy density will in general not be negligible at $t_{\rm r}$ and the universe is therefore not fully radiation dominated. However, decreasing the pressure decreases $C_{\rm c}$ and using the threshold $C_{\rm c}$ for radiation domination we should be estimating the PBH abundance from below. In any case, our results should be regarded as order of magnitude estimates both due to the use of Eq.~\eqref{eq:Mpbh} and due to the monopole truncation Eq.~\eqref{eq:chir}. 

The probability that a spherical overdensity coarse grained over the comoving radius $r$ collapses into a PBH upon horizon entry equals the probability that $C$ exceeds the threshold $C_{\rm c}$.  The contribution of the PBHs to the total energy density at the collapse time can then be written as 
\baq
\label{eq:beta}
\beta(r) \equiv \frac{\rho_{M(r)}}{\rho}\left.\rule{0pt}{3 ex}\right|_{t_{\rm r}}&=&\int_{C_{\rm c}}^{\frac{f(w)}{2}}\diffd C\ \frac{M(C,r)}{M_{\rm H}(r)}P_C(C,r)\\\nonumber
&=&
\int_{C_{l,\rm c}}^{f(w)}\diffd C_l\ K \left(C_l(r) - \frac{1}{2f(w)}C_l(r)^2 - C_{\rm c}\right)^{\gamma}P_{C_l}(C_l,r)~,
\eaq
where $C_{l,\rm c}= f(w)(1-\sqrt{1-2 C_{\rm c}/f(w)})$,  we used $\diffd C P_C(C,r) = \diffd C_l P_{C_l}(C_l,r)$,  and $P_{C_l}(C_l,r)$ is given by Eq.~\eqref{eq:fullpdf}. The upper limit $C = f(w)/2$ is the largest fluctuation amplitude that forms a type I overdensity for which the areal radius is a monotonic function of the coordinate $r$~\cite{Kopp:2010sh}. 

The fraction of the present day dark matter energy density constituted by the PBHs reads  
\beq
\label{eq:fpbh(r)}
f_{\rm PBH}(r) \equiv \frac{\rho_{M(r)}}{\rho_{\rm DM}}\left.\rule{0pt}{3 ex}\right|_{t_{0}} =  \frac{\Omega_{{\rm m},0}}{\Omega_{{\rm DM},0}} \frac{1}{k_{\rm eq} r}\beta(r)~,
\eeq 
where $k_{\rm eq} = (a H)_{\rm eq}$ at the matter radiation equality, and we have omitted the ${\cal O}(1)$ factor $(g_{*}(t_{\rm r})/g_{*}(t_{\rm eq}))^{-1/6}$ from the effective number of relativistic degrees of freedom.  
This can be recast as  
 \beq
f_{\rm PBH}(r) = \int {\rm d}{\rm ln} M f(M)~,
\eeq
with the mass distribution function $f(M)$ given by 
\beq
\label{eq:f(M)}
f(M) = \frac{\Omega_{{\rm m},0}}{\Omega_{{\rm DM},0}} \frac{1}{k_{\rm eq} r} 
\frac{\left(C_l-\frac{1}{2f(w)}C_l^2 -C_{\rm c}\right)^{\gamma+1}}{\gamma\left(1-\frac{1}{f(w)}C_l\right)}P_{C_l}(C_l,r)~,
\eeq
where $C_{l}\equiv C_{l}(M) = f(w)\left(1-\sqrt{1-[2/f(w)](C_{\rm c}+[M/(KM_{\rm H})]^{1/\gamma})}\right)$, as obtained from Eq.~\eqref{eq:Mpbh}.

\section{Results}
\label{sec:results}

It is straightforward to compute the variances Eq.~\eqref{eq:Sigmacomp} using Eq.~\eqref{eq:Pchipowerlaw} and numerically perform the integral in Eq.~\eqref{eq:fullpdf} to find the probability distribution $P_{C_l}(C_l,r)$.  Using Eqs.~\eqref{eq:beta} and \eqref{eq:fpbh(r)} we then get $f_{\rm PBH}(r)$ as a function of the coarse-graining scale $r$.

\begin{figure}[H]
\centering
\includegraphics[width=0.5\textwidth]{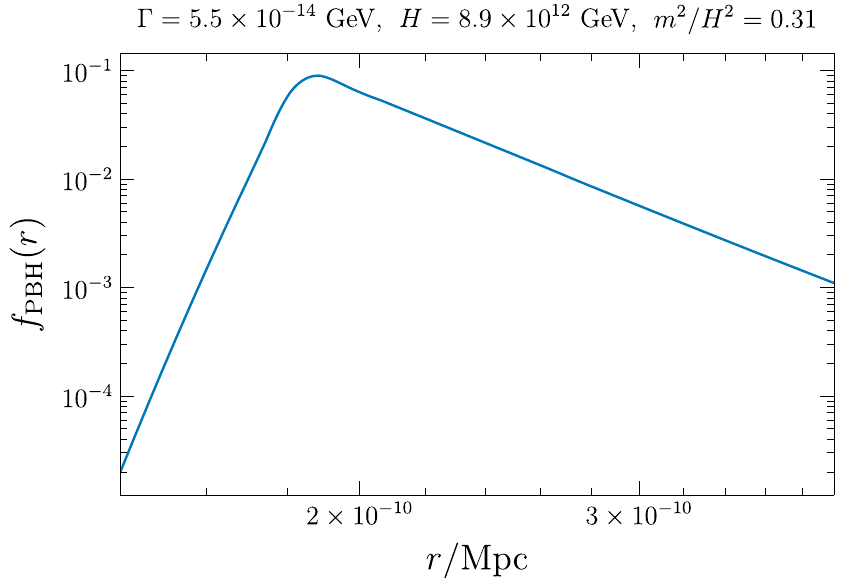}\includegraphics[width=0.5\textwidth]{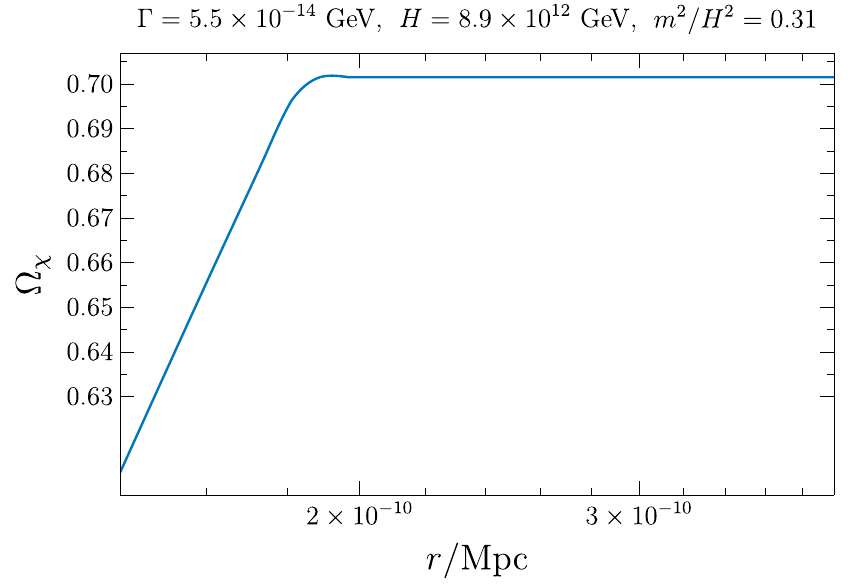}
\caption{Left panel: The fraction of the dark matter abundance in PBHs $f_{\rm PBH}$ as function of the coarse-graining scale $r$. Right panel: The curvaton density parameter $\Omega_{\chi}$ evaluated at $a(t_{\rm r}) H(t_{\rm r}) r = 1$ shown as a function of the coarse-graining scale $r$. Both panels are computed setting $\Gamma = 5.5\times 10^{-14}$ GeV, $H=8.9\times 10^{12}$ GeV, and $m^2/H^2 = 0.31$.}       
\label{fig:fovsrplot}
\end{figure}

Figure~\ref{fig:fovsrplot} illustrates the typical shape of the function $f_{\rm PBH}(r)$ and the curvaton density parameter $\Omega_{\chi}$ at the horizon crossing of $r$, which enters in the computation via Eq.~\eqref{eq:fullpdf}. The abundance $f_{\rm PBH}(r)$ has a clearly peaked structure although the curvaton spectrum Eq.~\eqref{eq:Pchipowerlaw} is of pure power law form. This is a generic feature in the setup and it arises from an interplay of two opposite effects.  First, increasing the coarse-graining scale $r$ makes the variance $\sigma^2_{\chi\chi}(r)$ smaller,  see Eq.~\eqref{eq:Sigmacomp}, and therefore suppresses the probability of large $\chi(r)$ values that can source PBHs.  Second, the curvature perturbation $\zeta$ and $\Omega_{\chi}$ keep growing in time until the curvaton decay at $H(t_{\rm dec}) = \Gamma$. For $t_{\rm r} < (t_{\rm dec})$,  increasing $r$ corresponds to later horizon crossing times $t_{\rm r}$, making $\zeta(t_{\rm r})$ larger and enhancing the probability for PBH formation. This effect dominates to the left of the $f_{\rm PBH}(r)$ peak in Fig.~\ref{fig:fovsrplot}, and  the peak corresponds to $t_{\rm r}=t_{\rm dec}$. To the right of the peak, $\zeta$ stays constant as $t_{\rm r}>t_{\rm dec}$. In this region, increasing $r$ only acts to decrease $\sigma^2_{\chi\chi}(r)$ and therefore $f_{\rm PBH}(r)$ starts to decrease. In the following, we will choose the coarse-graining scale $r$ equal to the peak scale by setting  $r=r_{\rm dec} \equiv 1/(a(t_{\rm dec}) H(t_{\rm dec}))$, and define $f_{\rm PBH} = f_{\rm PBH}(r_{\rm dec})$.

\begin{figure}[H]
\centering
\includegraphics[width=0.464 \textwidth]{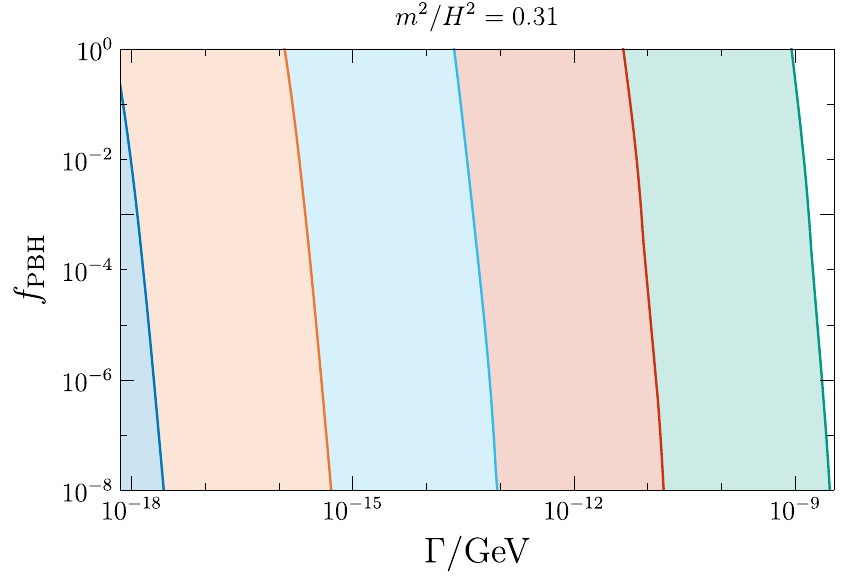}\includegraphics[width=0.536 \textwidth]{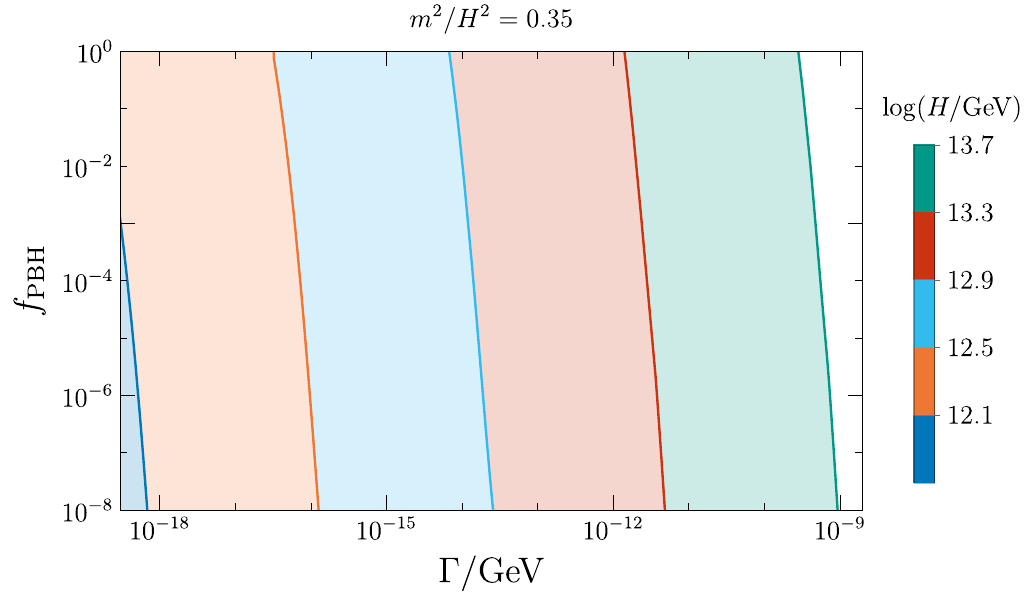}
\caption{PBH abundance $f_{\rm PBH}$ as a function of the curvaton decay rate $\Gamma$ and the inflationary Hubble scale $H$. Computed for $m^2/H^2 = 0.31$ (left) and 0.35 (right). }        
\label{fig:fvshplot}
\end{figure}

\begin{figure}[H]
\centering
\includegraphics[width=0.8 \textwidth]{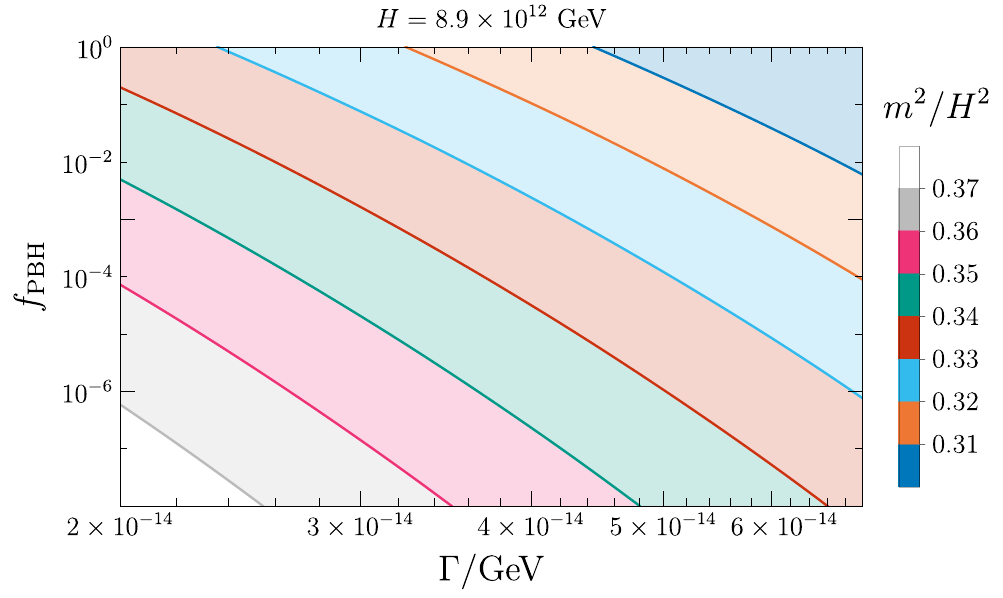}
\caption{PBH abundance $f_{\rm PBH}$ as a function of the curvaton decay rate $\Gamma$ and the curvaton mass parameter $m^2/H^2$. Computed for $H = 8.9\times 10^{12}$ GeV.}
\label{fig:fvsmplot}
\end{figure}

Figures~\ref{fig:fvshplot} and~\ref{fig:fvsmplot} illustrate the behaviour of $f_{\rm PBH}$ as a function of the decay rate $\Gamma$, the inflationary Hubble scale $H$, and the curvaton mass parameter $m^2/H^2$. Varying the decay rate $\Gamma$ alters the curvaton density parameter $\Omega_{\chi}$ at $t_{\rm r} = t_{\rm dec}$. As seen in Fig.~\ref{fig:fovsrplot}, $f_{\rm PBH}$ is a strongly dependent function of $\Omega_{\chi}$. Increasing $\Gamma$ moves the decay to earlier times and decreases $\Omega_{\chi}$ for fixed $H,m^2/H^2$ values. This explains the strong  decrease of $f_{\rm PBH}$ as a function of $\Gamma$ in Figs.~\ref{fig:fvshplot} and~\ref{fig:fvsmplot}. The figures further show that, for a fixed $\Gamma$, increasing $m^2/H^2$ or decreasing $H$ causes $f_{\rm PBH}$ to decrease. The former is because larger $m^2/H^2$ makes the curvaton spectrum Eq.~\eqref{eq:Pchipowerlaw} more blue-tilted, decreasing the variance $\sigma_{\chi\chi}^2(r)$, Eq.~\eqref{eq:Sigmacomp}, and therefore suppressing the probability for large $\chi(r)$ values. The latter is because the mean curvaton energy at the end of inflation $\langle\rho_{\chi}\rangle\propto  m^2 \langle \chi^2\rangle \propto H^4$~\cite{Starobinsky:1994bd}, and decreasing $H$ therefore decreases $\Omega_{\chi}$ for fixed $\Gamma$ and $m^2/H^2$ values.

\begin{figure}[H]
\centering
\includegraphics[width=0.5 \textwidth]{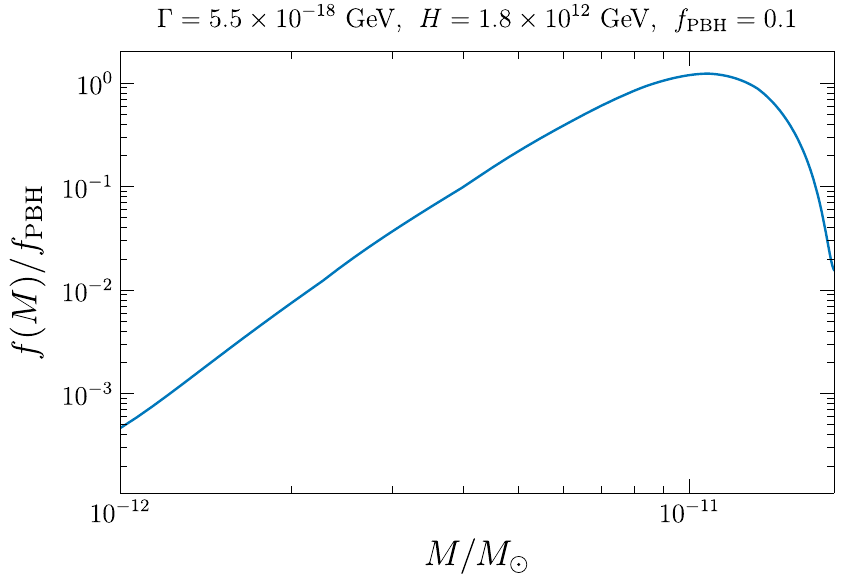}\includegraphics[width=0.5 \textwidth]{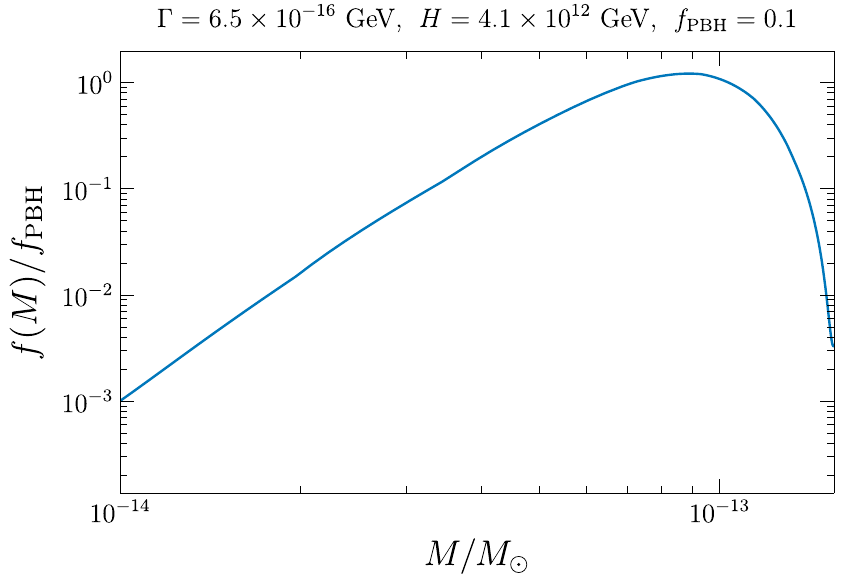}
\includegraphics[width=0.5 \textwidth]{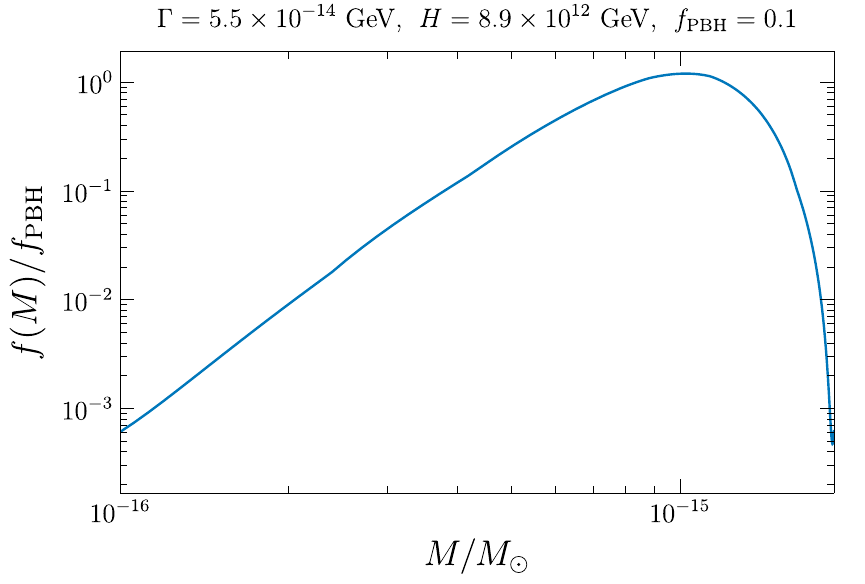}
\caption{The PBH mass distribution $f(M)$ normalised by the PBH abundance $f_{\rm PBH}$. In all figures $m^2/H^2=0.31$. The first panel shows the results for $\Gamma=5.5\times 10^{-18}$ GeV, $H\approx 1.78\times 10^{12}$ GeV, the second panel for 
$\Gamma=6.5\times 10^{-16}$ GeV, $H\approx4.08\times 10^{12}$ GeV, and the third panel for $\Gamma=5.5\times 10^{-14}$ GeV, $H\approx8.85\times 10^{12}$ GeV. In all cases $f_{\rm PBH} \approx 0.10$. }        
\label{fig:fmplot}
\end{figure}

The observational constraints on the PBH abundance $f_{\rm PBH}$ depend on the PBH mass which, setting $r=r_{\rm dec}$ in Eqs.~\eqref{eq:Mpbh} and \eqref{eq:MH}, in our setup is parametrically proportional to 
\beq
M_{\rm H} = 4 \pi M_{\rm P}^2 \Gamma^{-1} \approx 1.3\times 10^{-15} M_{\odot} \left(\frac{\Gamma}{5 \times 10^{-14}\;{\rm GeV}}\right)^{-1}~.
\eeq
Black hole evaporation sets very strong constraints for PBHs with $M\lesssim 10^{-16}  M_{\odot}$. Constraints for $M\gtrsim 10^{-10} M_{\odot}$ PBHs from various different systems range downwards from $f_{\rm PBH} = {\cal O}(10^{-2})$ depending on the mass~\cite{Carr:2020gox}. In the asteroid mass window $10^{-16} M_{\odot}\lesssim M \lesssim 10^{-10} M_{\odot}$, the constraints are subject to significant uncertainties and it can be possible to have $f_{\rm PBH} = 1$ in this window~\cite{Carr:2020gox}. Interestingly, asteroid mass PBHs can be efficiently produced in the curvaton scenario. This is demonstrated in Fig.~\ref{fig:fmplot} which shows the PBH mass spectra $f(M)$ computed using Eq.~\eqref{eq:f(M)} for $\Gamma = 5.5\times 10^{-18}$ GeV, $\Gamma = 6.5\times 10^{-16}$ GeV and $\Gamma = 5.5\times 10^{-14}$ GeV, with $m^2/H^2=0.31$, and $H$ chosen in each case such that $f_{\rm PBH} \approx 0.10$. The corresponding curvaton decay temperatures are $T_{\rm dec} \sim 2$ GeV for $\Gamma = 5.5\times 10^{-18}$ GeV, $T_{\rm dec} \sim 20$ GeV for $\Gamma = 6.5\times 10^{-16}$ GeV, and $T_{\rm dec} \sim 200$ GeV for $\Gamma = 5.5\times 10^{-14}$ GeV, approximating the curvaton decay into radiation and the thermalisation of the decay products as instant processes, and using the Standard Model $g_{*}(T)$. The respective mass spectra in Fig.~\ref{fig:fmplot} are peaked at $M\sim 10^{-11} M_{\odot}$, $M\sim 10^{-13} M_{\odot}$ and $M\sim 10^{-15} M_{\odot}$. In all three cases, the PBH abundance can be increased by slightly increasing $H$, see Fig.~\ref{fig:fvshplot}. For example, $f_{\rm PBH}\approx 1.0$ is obtained with $H\approx1.85\times 10^{12}$ GeV, $H\approx 4.25\times 10^{12}$ GeV, and  $H\approx 9.19\times 10^{12}$ GeV for $\Gamma = 5.5\times 10^{-18}$ GeV, $\Gamma = 6.5\times 10^{-16}$ GeV, and $\Gamma = 5.5\times 10^{-14}$ GeV, respectively.

Figures~\ref{fig:fvshplot},~\ref{fig:fvsmplot} and~\ref{fig:fmplot} represent the main results of this work. In particular, they show that the curvaton scenario can generate a significant dark matter fraction consisting of asteroid mass scale PBHs when the inflationary Hubble scale is large enough $H\gtrsim 10^{12}$ GeV, the curvaton mass parameter $m^2/H^2 \gtrsim 0.3$ and the decay rate falls in the window $10^{-18}\;{\rm GeV}\lesssim \Gamma \lesssim 10^{-13}$ GeV, corresponding to decay temperatures from slightly above the electroweak transition to around the QCD transition scale. 
If the curvaton decay occurs earlier,  $10^{-13}\; {\rm GeV}\lesssim \Gamma \lesssim 10^{-8}\; {\rm GeV}$, the scenario generates PBHs with  $M\lesssim 10^{-16}  M_{\odot}$ and, according to the dependencies shown in Figs.~\ref{fig:fvshplot} and~\ref{fig:fvsmplot}, the observational constraints on $f_{\rm PBH}$ constrains the viable parameter space for $H$ from above and for $m^2/H^2$ from below. For $\Gamma \gtrsim 10^{-8}\; {\rm GeV}$ there are no constraints as the PBH abundance is exponentially suppressed for any $H \lesssim 5.2 \times 10^{13}$ GeV compatible with the observational bound on the tensor-to-scalar ratio $r_{\rm T} < 0.044$~\cite{Tristram:2020wbi}, and $m^2/H^2$ in the range compatible with Eq.~\eqref{eq:Pzetachicond}\footnote{More precisely, for the maximal Hubble scale  $H = 5.2 \times 10^{13}$ GeV, Eq.~\eqref{eq:Pzetachicond} implies $m^2/H^2 \geqslant 0.29$, see Fig.~\ref{fig:m2overh2plot} in Appendix~\ref{app:spectrum_CMB}. The maximal PBH abundance is obtained for $m^2/H^2 = 0.29$ and we find $f_{\rm PBH} < 10^{-10}$ for $\Gamma > 7.45 \times 10^{-9}$ GeV.}. For $\Gamma \lesssim 10^{-18}\; {\rm GeV}$, the scenario generates PBHs with $M\gtrsim 10^{-10} M_{\odot}$ but in this region our numerical integration of Eq.~\eqref{eq:fullpdf} starts to become inaccurate for configurations leading to $f_{\rm PBH}\gtrsim 0.01$. We are therefore not able to perform a detailed study of the $\Gamma \lesssim 10^{-18}\; {\rm GeV}$ region in this work. 

\begin{figure}[H]
\centering
\includegraphics[width=0.7 \textwidth]{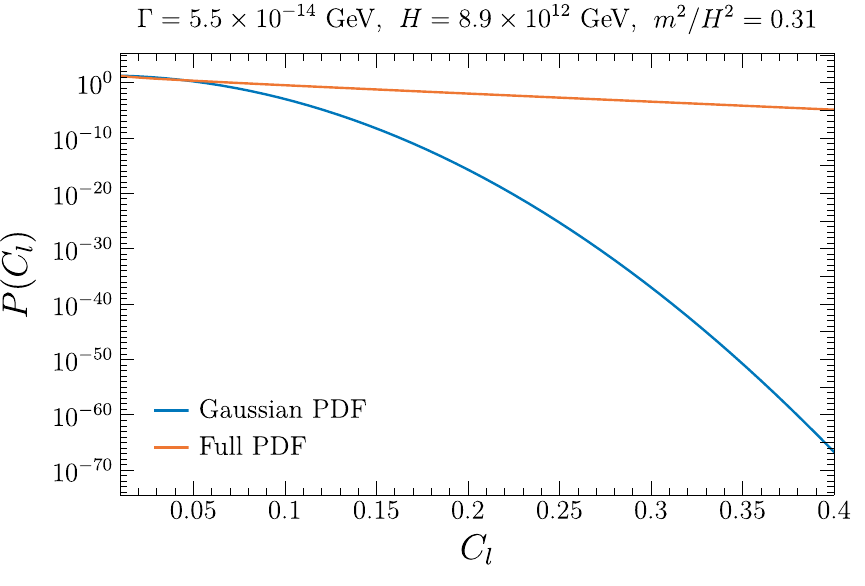}
\caption{The full probability distribution function $P_{C_l}(C_l,r_{\rm dec})$ and a Gaussian distribution with the same variance $\langle C_l^2\rangle \approx 5.1\times 10^{-4} $. Parameters chosen as in Fig.~\ref{fig:fovsrplot}.}        
\label{fig:pdfplot}
\end{figure}

Finally, Fig.~\ref{fig:pdfplot} depicts the probability distribution $P_{C_l}(C_l,r_{\rm dec})$ given by Eq.~\eqref{eq:fullpdf} for the same choice of parameters as in Fig.~\ref{fig:fovsrplot}, and in the third panel of Fig.~\ref{fig:fmplot}. For comparison, we also show a Gaussian distribution with the variance equal to the variance computed from the full distribution for this set of parameters, $\int \diffd C_l\ C_l^2 P_{C_l}(C_l,r_{\rm dec})\approx 5.1\times 10^{-4}$.  The full distribution deviates significantly from the Gaussian case and decays much slower as a function of $C_{l}$. The slowly decaying tail of $P_{C_l}(C_l,r_{\rm dec})$ is essential for the PBH formation in our setup. We recall, that the fully non-Gaussian form of $P_{C_l}(C_l,r_{\rm dec})$ follows from the vanishing mean of the curvaton field $\langle \chi \rangle =0$ which in turn is the equilibrium configuration during inflation for the $\chi^2$ potential. The curvature perturbation component $\zeta_{\chi}\propto {\rm ln} \rho_{\chi}/\langle \rho_{\chi}\rangle  ={\rm ln} \chi^2/\langle\chi^2\rangle$ has no leading Gaussian term and there is no suppression for fluctuations  $\chi^2/\langle\chi^2\rangle\sim 1$. Together with the non-linear form of Eq.~\eqref{eq:zetasolfull}, this gives rise to the strongly non-Gaussian distribution of $C_{l}$ seen in Fig.~\ref{fig:pdfplot}.

\section{Summary}
\label{sec:summary}

In this work we have investigated the PBH production in the mixed inflaton--curvaton scenario with a quadratic curvaton potential and a strongly blue-tilted curvaton spectrum, assuming the curvaton is in the de Sitter equilibrium $\langle \chi\rangle = 0$ during inflation. We require that the inflaton sourced Gaussian component dominates the spectrum of the curvature perturbation $\zeta$ on CMB scales and the curvaton sourced contribution is suppressed below the $2\times 10^{-11}$ level at the pivot scale $k_{*}=0.05\ {\rm Mpc}^{-1}$. This constrains the curvaton mass from below, for example for the inflationary Hubble scale $H = 10^{13}$ GeV, the curvaton mass must satisfy $m^2/H^2 \gtrsim 0.3$. On small scales, however, the curvaton sourced component of $\zeta$ can take large values leading to PBH formation.  

A key feature in this setup, is that the curvaton sourced part of $\zeta$ is strongly non-Gaussian. It contains no leading Gaussian term, unlike in scenarios with a non-vanishing curvaton mean value $|\langle \chi\rangle|\gg H$ (a specific initial condition during inflation) which have been studied previously in the PBH context \textit{e.g.}~in~\cite{Klimai:2012sf,Young:2013oia,Kawasaki:2012wr,Pi:2021dft,Meng:2022low}. We use the $\delta N$ formalism to obtain the full non-linear solution for $\zeta$ as a function of $\chi$ and, following~\cite{Gow:2022jfb}, include only the monopole term of the spherical harmonics series of $\chi({\bf x})$ when considering fluctuations relevant for PBHs. With this approximation, we compute the probability distribution for $C_{l}(r)$ which determines the compaction function $C(C_{l}(r))$, and where $r$ is the coarse-graining scale of perturbations. Comparing to a fiducial Gaussian distribution with the same variance $\langle C_{l}^2\rangle$, we find that the full distribution $P_{l}(C_{l},r)$ decreases exponentially slower as a function of $C_{l}$. We use $C_{\rm c} =0.55$ as the threshold for the PBH collapse~\cite{Musco:2020jjb} and model the PBH mass with parameters obtained for the power-law spectrum in~\cite{Musco:2008hv,Musco:2020jjb}.  

We find that the curvaton scenario with $\langle \chi \rangle = 0$ can lead to very efficient production of PBHs. In particular, the setup can generate asteroid mass PBHs, $10^{-16}M_{\odot}\lesssim M\lesssim 10^{-10}M_{\odot}$, with an abundance equal to the observed dark matter abundance, $f_{\rm PBH} = 1$, when the Hubble scale at the end of inflation $H\gtrsim 10^{12}$ GeV, the curvaton mass $m^2/H^2\gtrsim 0.3$ and the curvaton decay occurs in the window from slightly before the electroweak transition to around the QCD transition. If the curvaton decays before or after the aforementioned window, PBHs with masses below or above the asteroid mass range can be generated, respectively. The PBH abundance depends sensitively on $H$, $m^2/H^2$ and the curvaton decay rate $\Gamma$, and the observational bounds on $f_{\rm PBH}$ imply non-trivial constraints on these parameters when $\Gamma\lesssim 10^{-8}$ GeV. For larger values of $\Gamma$ the PBH production in the setup is exponentially suppressed.  

The main uncertainties in our results arise from the monopole truncation Eq.~\eqref{eq:chir} and the phenomenological choice of collapse parameters in Eq.~\eqref{eq:Mpbh}. Changing the collapse parameter values would affect $f_{\rm PBH}$ predicted for a fixed set of curvaton parameters. However, even order of magnitude changes of $f_{\rm PBH}$ are compensated by just slight changes of $H$, $m^2/H^2$ and $\Gamma$, as can be seen in Figs.~\ref{fig:fvshplot} and~\ref{fig:fvsmplot}. The error caused by the use of Eq.~\eqref{eq:chir} is harder to quantify but we expect that the very efficient PBH production from the strongly non-Gaussian $\zeta$ is a robust conclusion. 

\acknowledgments
AG is supported by the Science and Technology Facilities Council [grant numbers ST/S000550/1, ST/W001225/1]. For the purpose of open access, the authors have applied a Creative Commons Attribution (CC BY) licence to any Author Accepted Manuscript version arising. Supporting research data are available on reasonable request from the corresponding author.

\appendix

\newpage
\section{The spectrum of \texorpdfstring{$\zeta_{\chi}$}{ζ\textunderscore χ}}
\label{app:spectrum_zetachi}

We use the stochastic formalism~\cite{Starobinsky:1994bd} and the spectral expansion method~\cite{Starobinsky:1986fx,Markkanen:2019kpv,Markkanen:2020bfc} to compute the infrared spectrum of $\zeta_{\chi}$ in our setup where $\langle \chi \rangle = 0$ and Eq.~\eqref{eq:zetachi} cannot be expanded in small perturbations around a mean field.  For the quadratic curvaton potential (\ref{Vchi}),  the joint equal time two-point distribution of $\chi(t,{\bf r})$ in de Sitter equilibrium is given by the spectral sum~\cite{Starobinsky:1986fx,Markkanen:2019kpv,Markkanen:2020bfc}
\beq
\label{eq:rho2}
\rho_{2}(\chi,{\bf r},t;\chi',{\bf r}',t) =\frac{m^2}{H^{4}}\psi_{0}(x)\psi_{0}(x')\sum_{n=0}^{\infty}\psi_n(x)\psi_n(x')\left(a(t)H|{\bf r}-{\bf r}^{\prime}|\right)^{-2\Lambda_n/H}\;,
\eeq
where 
\beq
x=\frac{m\chi}{H^2}\;,\quad \Lambda_n =\frac{nm^{2}}{3H}\;,\quad
\psi_{n}(x)=\frac{1}{\sqrt{n!2^{n}}}\left(\frac{4\pi}{3}\right)^{1/4}e^{-2\pi^{2}x^{2}/3}H_{n}\left(\frac{2\pi x}{\sqrt{3}}\right)\;,
\eeq
and $H_{n}(x) = (-1)^{n}e^{x^{2}/2}\frac{\diffd^{n}}{\diffd x^{n}}e^{-x^{2}/2}$ are the Hermite polynomials.  The eigenfunctions $\psi_{n}(x)$ are orthonormal, 
\beq
    \int_{-\infty}^{\infty}\diffd x\ \psi_{m}(x)\psi_{n}(x) = \delta_{mn}\;.
\eeq
Using that $\langle \chi^2\rangle = {3H^{4}}/({8\pi^{2}m^{2}})$,  we have from Eq.~\eqref{eq:zetachi}  $\zeta_{\chi} =(1/3) {\rm ln} (8\pi^3x^2/3)$,  and the connected part of its two-point function can be written as 
\baq
    \langle{\zeta}_{\chi}({\bf r}){\zeta}_{\chi}({\bf r}')\rangle_{\rm c} &=&\int_{-\infty}^{\infty}\diffd\chi\int_{-\infty}^{\infty}\diffd\chi'\ {\zeta}(\chi){\zeta}(\chi')\rho_{2}(\chi,{\bf r},t;\chi^{\prime},{\bf r}',t)\nonumber\\
    &=&\sum_{n=1}^{\infty}\left(\int_{-\infty}^{\infty}\diffd x\ \frac{1}{3}\psi_{0}(x)\psi_{n}(x)\ln{x}^{2}\right)^{2}\left(a(t)H\vert\textbf{r}-\textbf{r}^{\prime}\vert\right)^{-\frac{2 n m^{2}}{3H^{2}}}\;.
\eaq
Truncating the series at the leading order we get 
\beq
    \langle{\zeta}_{\chi}({\bf r}){\zeta}_{\chi}({\bf r}')\rangle_{\rm c} = \frac{2}{9}\left(aH |{\bf r}-{\bf r}'|\right)^{-\frac{4 m^{2}}{3H^{2}}}+\mathcal{O}\left[\left(aH |{\bf r}-{\bf r}'|\right)^{-\frac{8 m^{2}}{3H^{2}}}\right]\;.
\eeq
The corresponding power spectrum is given by 
\baq
\label{eq:appaPzetachi}
{\cal P}_{\zeta_{\chi}} &=& \frac{k^{3}}{2\pi^{2}}\int \diffd^{3}r\  e^{-i\textbf{k}\cdot\textbf{r}} \langle{\zeta}_{\chi}({\bf r}){\zeta}_{\chi}({\bf r}')\rangle_{\rm c}\nonumber\\
&=& \frac{2^{3-\frac{4m^{2}}{3H^{2}}}}{9\sqrt{\pi}}\frac{\Gamma\left(\frac{3}{2}-\frac{2m^{2}}{3H^{2}}\right)}{\Gamma\left(\frac{2m^{2}}{3H^{2}}\right)}\left(\frac{k}{aH}\right)^{\frac{4m^{2}}{3H^{2}}}\;.
\eaq
The existence of the Fourier transform requires $4m^2 < 9 H^2$ which we assume here.  

\begin{figure}[H]
\centering
\includegraphics[width=0.7 \textwidth]{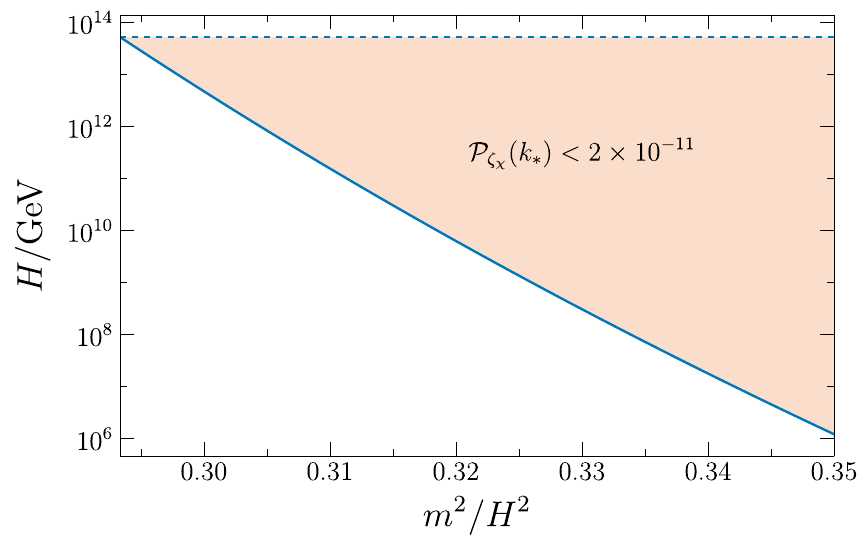}
\caption{Above the solid line ${\cal P}_{\zeta_{\chi}}(k_*) < 2 \times 10^{-11}$. The dashed line shows the maximum inflationary Hubble scale $H \approx 5.2 \times 10^{13}\;{\rm GeV}$ consistent with the observational bound on the tensor-to-scalar ratio $r_{\rm T} < 0.044$~\cite{Tristram:2020wbi}. The shaded orange area between the solid and dashed lines depicts the phenomenologically viable region where ${\cal P}_{\zeta_{\chi}}(k_*) < 2 \times 10^{-11}$.}
\label{fig:m2overh2plot}        
\end{figure}

Assuming an instant transition from de Sitter inflation to radiation dominated epoch,  and  approximating the universe as radiation dominated until the curvaton decay\footnote{This is strictly valid for $\Omega_{\chi} \ll 1$ but suffices here because we only consider cases where the curvaton decays before it fully dominates the universe, and therefore the period when $\Omega_{\chi} \ll1$ may not hold is short.}, we can write   
\beq
\label{eq:appakoverkend}
    \frac{k}{aH} = 6\times 10^{-24} \frac{k}{0.05\ {\rm Mpc}^{-1}}\frac{10^{15} {\rm GeV}}{\rho^{1/4}}~,
\eeq
where $\rho = 3 H^2 M_{\rm P}^2$ denotes the energy density during inflation. Using Eqs.~\eqref{eq:appaPzetachi} and \eqref{eq:appakoverkend} we can directly solve for the $m^2/H^2$ range where the condition Eq.~\eqref{eq:Pzetachicond} is satisfied, \textit{i.e.} ${\cal P}_{\zeta_{\chi}}(k_*) < 2 \times 10^{-11}$. The result is shown in Fig.~\ref{fig:m2overh2plot} where the shaded orange area marks the region where Eq.~\eqref{eq:Pzetachicond} holds. 

We will also need the two point functions of the powers $\zeta^{\ell}_{\chi}$ in the analysis below. These can be directly computed using the spectral expansion expression 
\baq
    \langle{\zeta}_{\chi}^{{\ell}}({\bf r}){\zeta}_{\chi}^{{\ell}'}({\bf r}')\rangle_{\rm c} &=&\int_{-\infty}^{\infty}\diffd\chi\int_{-\infty}^{\infty}\diffd\chi'\ {\zeta}^{\ell}(\chi){\zeta}^{\ell'}(\chi')\rho_{2}(\chi,{\bf r},t;\chi^{\prime},{\bf r}',t)\\\nonumber
    &=&\sum_{n=1}^{\infty}\sum_{n'=1}^{\infty}\left(\int_{-\infty}^{\infty}\hspace{-7pt}\diffd x\ \frac{1}{3}\psi_{0}(x)\psi_{n}(x)\ln{x}^{2}\right)^{{\ell}}
    \left(\int_{-\infty}^{\infty}\hspace{-7pt}\diffd x'\ \frac{1}{3}\psi_{0}(x')\psi_{n}(x')\ln{x'}^{2}\right)^{{\ell}'}\times
    \\\nonumber
    &&
    \left(a(t)H\vert\textbf{r}-\textbf{r}^{\prime}\vert\right)^{-\frac{2 (n+n')m^{2}}{3H^{2}}}~.
\eaq
Truncating the series again at the leading order we obtain 
\beq
\label{A:zetachinpoint}
    \langle{\zeta}_{\chi}^{{\ell}}({\bf r}){\zeta}_{\chi}^{{\ell}'}({\bf r}')\rangle_{\rm c} =\left(\frac{\sqrt{2}}{3}\right)^{\ell+\ell'}\left(aH |{\bf r}-{\bf r}'|\right)^{-\frac{4 m^{2}}{3H^{2}}}+\mathcal{O}\left[\left(aH |{\bf r}-{\bf r}'|\right)^{-\frac{8 m^{2}}{3H^{2}}}\right]~.
\eeq

\section{The spectrum of \texorpdfstring{$\zeta$}{ζ} on CMB scales}
\label{app:spectrum_CMB}

We express the full solution Eq.~\eqref{eq:zetasolfull} for $\zeta$ in the form   
\beq
\zeta = \zeta_{\rm r} + Z(\zeta_{\chi},\zeta_{\rm r}) 
~,
\eeq
where the explicit expression for $Z(\zeta_{\chi},\zeta_{\rm r})$ can be read off from Eq.~\eqref{eq:zetasolfull}. 
The connected part of the two point function of $\zeta$ is given by  
\beq
\label{B:zeta2pointfull}
\langle\zeta({\bf r})\zeta({\bf r}')\rangle_{\rm c}=\langle\zeta_{\rm r}({\bf r})\zeta_{\rm r}({\bf r}')\rangle_{\rm c}+2\langle Z({\bf r})\zeta_{\rm r}({\bf r}')\rangle_{\rm c}+\langle Z({\bf r})Z({\bf r}')\rangle_{\rm c}~.
\eeq
Expanding $Z(\zeta_{\chi}-\zeta_{\rm r})$ around $\zeta_{\rm r} = 0$,  the last two terms can be written as 
 \baq
\label{B:zeta2pointpart1}
 \langle Z({\bf r})\zeta_{\rm r}({\bf r}')\rangle_{\rm c}&=&\sum_{n=0}^{\infty}\frac{1}{(2n+1)!} \langle Z^{(2n+1)}(\zeta_{\chi})\rangle \langle\zeta_{\rm r}({\bf r})\zeta_{\rm r}^{2n+1}
 ({\bf r}')\rangle_{\rm c}\\
 \label{B:zeta2pointpart2}
\langle Z({\bf r})Z({\bf r}')\rangle_{\rm c} &=&\sum_{n=1}^{\infty}\sum_{n'=1}^{\infty}\frac{1}{n!n'!} \langle Z^{(n)}(\zeta_{\chi})\rangle \langle Z^{(n')}(\zeta_{\chi})\rangle \langle\zeta_{\rm r}^n({\bf r})\zeta_{\rm r}^{n'}({\bf r}')\rangle_{\rm c}\\\nonumber
&&+{~\rm terms~involving~} \langle\zeta_{\chi}^n({\bf r})\zeta_{\chi}^{n'} ({\bf r}')\rangle_{\rm c}
~.
 \eaq

We assume the curvature perturbation of the radiation component is sourced by the inflaton,  and is Gaussian distributed with a nearly scale invariant spectrum 
\beq
 \label{B:zetar2point}
{\cal P}_{\zeta_{\rm r}}(k) = A_{\rm r}\left(\frac{k}{k_*}\right)^{n_{\rm r}-1}~,\qquad A_{\rm r} \gg P_{\zeta_{\chi}}(k_*)~, 
\eeq
and require that the curvaton component is suppressed on CMB scales such that $P_{\zeta_{\chi}}(k_*) < 2\times 10^{-11}$ according to Eq.~\eqref{eq:Pzetachicond}.  Using Eq.~\eqref{A:zetachinpoint},  we then observe that on scales $k \lesssim k_*$ we can to leading precision drop all terms involving connected correlators of $\zeta_{\chi}^{n}$ in Eq.~\eqref{B:zeta2pointpart2}.  On scales $k \lesssim k_*$,  the leading contribution to the two point function Eq.~\eqref{B:zeta2pointfull} then reads 
\beq
\label{B:zeta2pointleading}
\langle\zeta({\bf r})\zeta({\bf r}')\rangle_{\rm c}\approx\langle\zeta_{\rm r}({\bf r})\zeta_{\rm r}({\bf r}')\rangle_{\rm c}\left(1+\left\langle Z^{(1)}(\zeta_{\chi})\right\rangle\right)^2~.
\eeq
The non-linear function $\langle Z^{(1)}(\zeta_{\chi})\rangle$ involves contributions from closed $\zeta_{\chi}$ loops attached to the points ${\bf r}$ or ${\bf r}'$.  Since $\langle Z^{(1)}(\zeta_{\chi})\rangle$ has no dependence on spatial coordinates,  the spectrum of the curvature perturbation for $k \lesssim k_*$ is simply given by\
\beq
 \label{B:Pzetaleading}
{\cal P}_{\zeta}(k) = A \left(\frac{k}{k_*}\right)^{n_{\rm r}}~,\qquad A=A_{\rm r}\left(1+\left\langle Z^{(1)}(\zeta_{\chi})\right\rangle\right)^2 ~,\qquad k\lesssim k_*~.
\eeq

\begin{figure}[H]
\centering
\includegraphics[width=0.7 \textwidth]{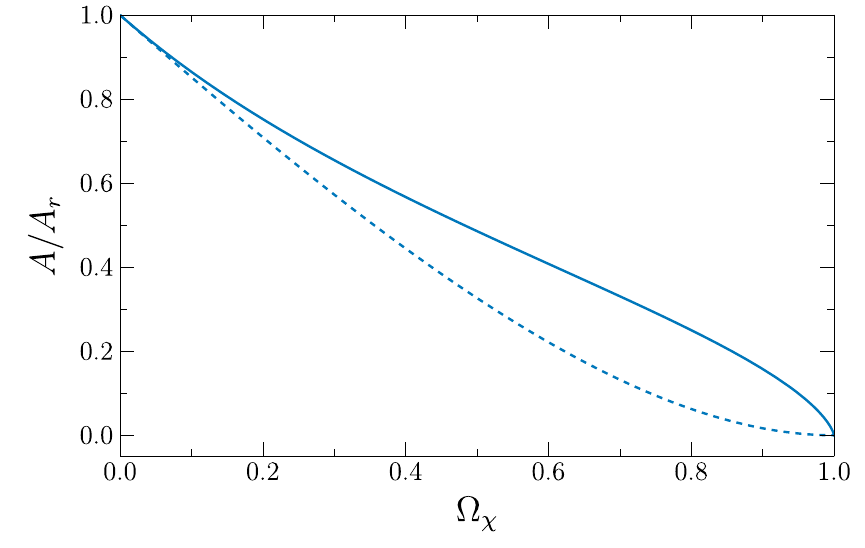}
\caption{The solid line shows $A/A_{\rm r}$ defined in Eq.~\eqref{B:Pzetaleading} as a function of the curvaton density parameter $\Omega_{\chi}$. The dashed line shows the corresponding quantity computed in linear perturbation theory.}
\label{fig:AoverArplot}        
\end{figure}

Figure~\ref{fig:AoverArplot} shows the ratio $A/A_{\rm r}$ as function of $\Omega_{\chi}$. For comparison, we have also plotted the corresponding linear perturbation theory result $A/A_{\rm r} = (4(1-\Omega_{\chi})/(4-\Omega))^2$ obtained from $\zeta = 4\zeta_{\rm r}(1-\Omega_{\chi})/(4-\Omega)+3\zeta_{\chi}\Omega_{\chi}/(4-\Omega)$, see \textit{e.g.}~\cite{Langlois:2004nn}, when ${\cal P}_{\zeta_{\chi}}(k) \ll {\cal P}_{\zeta_{\rm r}}(k)$. As seen in Fig.~\ref{fig:AoverArplot}, for $\Omega_{\chi}\lesssim 0.9$ we have $0.2 \lesssim A/A_{\rm r}\leqslant 1 $, indicating that the observed CMB spectrum amplitude ${\cal P}_{\zeta}(k_*)= A  = 2.10 \times 10^{-9}$ is obtained by adjusting the inflaton sector to give $A_{\rm r}$ in the range $5 \gtrsim  A_{\rm r}\geqslant 1 $, correspondingly. From Eqs.~\eqref{A:zetachinpoint} and \eqref{B:zeta2pointfull}--\eqref{B:zeta2pointpart2} we further see that when Eq.~\eqref{eq:Pzetachicond} holds,  the contributions from $\zeta_{\chi}$ to the spectral index $n_{\rm s}-1= \diffd{\rm ln} P_{\zeta}/\diffd{\rm ln} k$ are parametrically suppressed to ${\cal O}(0.01)$.  The measured spectral index $n_{\rm s}= 0.965$ at $k = k_{*}$ should then be obtainable with slow roll inflationary models having $\epsilon = {\cal O}(0.01)$ which can be adjusted to give the correct $n_{\rm s}$.  We thus conclude that when Eq.~\eqref{eq:Pzetachicond} holds, we obtain inflaton dominated, Gaussian and nearly scale-invariant perturbations on CMB scales.
 

\bibliography{Curvaton_non-Gaussianity_paper}

\providecommand{\href}[2]{#2}\begingroup\raggedright\begin{thebibliography}{10}

\bibitem{Zeldovich:1967lct}
{\relax Ya}.~B. Zel'dovich and I.~D. Novikov, \emph{{The Hypothesis of Cores
  Retarded during Expansion and the Hot Cosmological Model}},
  \href{https://ui.adsabs.harvard.edu/abs/1967SvA....10..602Z/abstract}{\emph{Soviet
  Astronomy} {\bfseries 10}{\bfseries (4)} (1967) 602}.

\bibitem{Hawking:1971ei}
S.~Hawking, \emph{{Gravitationally Collapsed Objects of Very Low Mass}},
  \href{https://doi.org/10.1093/mnras/152.1.75}{\emph{Monthly Notices of the
  Royal Astronomical Society} {\bfseries 152}{\bfseries (1)} (1971) 75}.

\bibitem{Chapline:1975ojl}
G.~F. Chapline, \emph{Cosmological effects of primordial black holes},
  \href{https://doi.org/10.1038/253251a0}{\emph{Nature} {\bfseries
  253}{\bfseries (5489)} (1975) 251}.

\bibitem{Carr:2020gox}
B.~Carr, K.~Kohri, Y.~Sendouda and J.~Yokoyama, \emph{Constraints on primordial
  black holes}, \href{https://doi.org/10.1088/1361-6633/ac1e31}{\emph{Reports
  on Progress in Physics} {\bfseries 84}{\bfseries (11)} (2021) 116902}
  [\href{https://arxiv.org/abs/2002.12778}{{\ttfamily 2002.12778}}].

\bibitem{Bird:2016dcv}
S.~Bird \textit{et~al.}, \emph{{Did LIGO detect dark matter?}},
  \href{https://doi.org/10.1103/PhysRevLett.116.201301}{\emph{Physical Review
  Letters} {\bfseries 116}{\bfseries (20)} (2016) 201301}
  [\href{https://arxiv.org/abs/1603.00464}{{\ttfamily 1603.00464}}].

\bibitem{Clesse:2016vqa}
S.~Clesse and J.~García-Bellido, \emph{{The clustering of massive Primordial
  Black Holes as Dark Matter: measuring their mass distribution with Advanced
  LIGO}}, \href{https://doi.org/10.1016/j.dark.2016.10.002}{\emph{Physics of
  the Dark Universe} {\bfseries 15} (2017) 142}
  [\href{https://arxiv.org/abs/1603.05234}{{\ttfamily 1603.05234}}].

\bibitem{Sasaki:2016jop}
M.~Sasaki, T.~Suyama, T.~Tanaka and S.~Yokoyama, \emph{{Primordial Black Hole
  Scenario for the Gravitational-Wave Event GW150914}},
  \href{https://doi.org/10.1103/PhysRevLett.117.061101}{\emph{Physical Review
  Letters} {\bfseries 117}{\bfseries (6)} (2016) 061101}
  [\href{https://arxiv.org/abs/1603.08338}{{\ttfamily 1603.08338}}].

\bibitem{Gow:2019pok}
A.~D. Gow, C.~T. Byrnes, A.~Hall and J.~A. Peacock, \emph{{Primordial black
  hole merger rates: distributions for multiple LIGO observables}},
  \href{https://doi.org/10.1088/1475-7516/2020/01/031}{\emph{Journal of
  Cosmology and Astroparticle Physics} {\bfseries 2020}{\bfseries (01)} (2020)
  031} [\href{https://arxiv.org/abs/1911.12685}{{\ttfamily 1911.12685}}].

\bibitem{Franciolini:2021tla}
G.~Franciolini \textit{et~al.}, \emph{{Searching for a subpopulation of
  primordial black holes in LIGO-Virgo gravitational-wave data}},
  \href{https://doi.org/10.1103/PhysRevD.105.083526}{\emph{Physical Review D}
  {\bfseries 105}{\bfseries (8)} (2022) 083526}
  [\href{https://arxiv.org/abs/2105.03349}{{\ttfamily 2105.03349}}].

\bibitem{Ivanov:1994pa}
P.~Ivanov, P.~Naselsky and I.~Novikov, \emph{Inflation and primordial black
  holes as dark matter},
  \href{https://doi.org/10.1103/PhysRevD.50.7173}{\emph{Physical Review D}
  {\bfseries 50}{\bfseries (12)} (1994) 7173}.

\bibitem{Garcia-Bellido:1996mdl}
J.~García-Bellido, A.~Linde and D.~Wands, \emph{Density perturbations and
  black hole formation in hybrid inflation},
  \href{https://doi.org/10.1103/PhysRevD.54.6040}{\emph{Physical Review D}
  {\bfseries 54}{\bfseries (10)} (1996) 6040}
  [\href{https://arxiv.org/abs/astro-ph/9605094}{{\ttfamily
  astro-ph/9605094}}].

\bibitem{Ivanov:1997ia}
P.~Ivanov, \emph{Nonlinear metric perturbations and production of primordial
  black holes}, \href{https://doi.org/10.1103/PhysRevD.57.7145}{\emph{Physical
  Review D} {\bfseries 57}{\bfseries (12)} (1998) 7145}
  [\href{https://arxiv.org/abs/astro-ph/9708224}{{\ttfamily
  astro-ph/9708224}}].

\bibitem{Leach:2000ea}
S.~M. Leach, I.~J. Grivell and A.~R. Liddle, \emph{Black hole constraints on
  the running mass inflation model},
  \href{https://doi.org/10.1103/PhysRevD.62.043516}{\emph{Physical Review D}
  {\bfseries 62}{\bfseries (4)} (2000) 043516}
  [\href{https://arxiv.org/abs/astro-ph/0004296}{{\ttfamily
  astro-ph/0004296}}].

\bibitem{Drees:2011hb}
M.~Drees and E.~Erfani, \emph{Running-mass inflation model and primordial black
  holes}, \href{https://doi.org/10.1088/1475-7516/2011/04/005}{\emph{Journal of
  Cosmology and Astroparticle Physics} {\bfseries 2011}{\bfseries (04)} (2011)
  005} [\href{https://arxiv.org/abs/1102.2340}{{\ttfamily 1102.2340}}].

\bibitem{Drees:2011yz}
M.~Drees and E.~Erfani, \emph{Running spectral index and formation of
  primordial black hole in single field inflation models},
  \href{https://doi.org/10.1088/1475-7516/2012/01/035}{\emph{Journal of
  Cosmology and Astroparticle Physics} {\bfseries 2012}{\bfseries (01)} (2012)
  035} [\href{https://arxiv.org/abs/1110.6052}{{\ttfamily 1110.6052}}].

\bibitem{Garcia-Bellido:2017mdw}
J.~García-Bellido and E.~Ruiz~Morales, \emph{Primordial black holes from
  single field models of inflation},
  \href{https://doi.org/10.1016/j.dark.2017.09.007}{\emph{Physics of the Dark
  Universe} {\bfseries 18} (2017) 47}
  [\href{https://arxiv.org/abs/1702.03901}{{\ttfamily 1702.03901}}].

\bibitem{Domcke:2017fix}
V.~Domcke, F.~Muia, M.~Pieroni and L.~T. Witkowski, \emph{{PBH dark matter from
  axion inflation}},
  \href{https://doi.org/10.1088/1475-7516/2017/07/048}{\emph{Journal of
  Cosmology and Astroparticle Physics} {\bfseries 2017}{\bfseries (07)} (2017)
  048} [\href{https://arxiv.org/abs/1704.03464}{{\ttfamily 1704.03464}}].

\bibitem{Kannike:2017bxn}
K.~Kannike, L.~Marzola, M.~Raidal and H.~Veermäe, \emph{Single field double
  inflation and primordial black holes},
  \href{https://doi.org/10.1088/1475-7516/2017/09/020}{\emph{Journal of
  Cosmology and Astroparticle Physics} {\bfseries 2017}{\bfseries (09)} (2017)
  020} [\href{https://arxiv.org/abs/1705.06225}{{\ttfamily 1705.06225}}].

\bibitem{Germani:2017bcs}
C.~Germani and T.~Prokopec, \emph{On primordial black holes from an inflection
  point}, \href{https://doi.org/10.1016/j.dark.2017.09.001}{\emph{Physics of
  the Dark Universe} {\bfseries 18} (2017) 6}
  [\href{https://arxiv.org/abs/1706.04226}{{\ttfamily 1706.04226}}].

\bibitem{Motohashi:2017kbs}
H.~Motohashi and W.~Hu, \emph{Primordial black holes and slow-roll violation},
  \href{https://doi.org/10.1103/PhysRevD.96.063503}{\emph{Physical Review D}
  {\bfseries 96}{\bfseries (6)} (2017) 063503}
  [\href{https://arxiv.org/abs/1706.06784}{{\ttfamily 1706.06784}}].

\bibitem{Ballesteros:2017fsr}
G.~Ballesteros and M.~Taoso, \emph{Primordial black hole dark matter from
  single field inflation},
  \href{https://doi.org/10.1103/PhysRevD.97.023501}{\emph{Physical Review D}
  {\bfseries 97}{\bfseries (2)} (2018) 023501}
  [\href{https://arxiv.org/abs/1709.05565}{{\ttfamily 1709.05565}}].

\bibitem{Hertzberg:2017dkh}
M.~P. Hertzberg and M.~Yamada, \emph{{Primordial Black Holes from Polynomial
  Potentials in Single Field Inflation}},
  \href{https://doi.org/10.1103/PhysRevD.97.083509}{\emph{Physical Review D}
  {\bfseries 97}{\bfseries (8)} (2018) 083509}
  [\href{https://arxiv.org/abs/1712.09750}{{\ttfamily 1712.09750}}].

\bibitem{Pi:2017gih}
S.~Pi, Y.-l. Zhang, Q.-G. Huang and M.~Sasaki, \emph{{Scalaron from
  $R^2$-gravity as a heavy field}},
  \href{https://doi.org/10.1088/1475-7516/2018/05/042}{\emph{Journal of
  Cosmology and Astroparticle Physics} {\bfseries 2018}{\bfseries (05)} (2018)
  042} [\href{https://arxiv.org/abs/1712.09896}{{\ttfamily 1712.09896}}].

\bibitem{Kohri:2018qtx}
K.~Kohri and T.~Terada, \emph{{Primordial black hole dark matter and LIGO/Virgo
  merger rate from inflation with running spectral indices: formation in the
  matter- and/or radiation-dominated universe}},
  \href{https://doi.org/10.1088/1361-6382/aaea18}{\emph{Classical and Quantum
  Gravity} {\bfseries 35}{\bfseries (23)} (2018) 235017}
  [\href{https://arxiv.org/abs/1802.06785}{{\ttfamily 1802.06785}}].

\bibitem{Biagetti:2018pjj}
M.~Biagetti, G.~Franciolini, A.~Kehagias and A.~Riotto, \emph{Primordial black
  holes from inflation and quantum diffusion},
  \href{https://doi.org/10.1088/1475-7516/2018/07/032}{\emph{Journal of
  Cosmology and Astroparticle Physics} {\bfseries 2018}{\bfseries (07)} (2018)
  032} [\href{https://arxiv.org/abs/1804.07124}{{\ttfamily 1804.07124}}].

\bibitem{Dalianis:2018frf}
I.~Dalianis, A.~Kehagias and G.~Tringas, \emph{Primordial black holes from
  $\alpha$-attractors},
  \href{https://doi.org/10.1088/1475-7516/2019/01/037}{\emph{Journal of
  Cosmology and Astroparticle Physics} {\bfseries 2019}{\bfseries (01)} (2019)
  037} [\href{https://arxiv.org/abs/1805.09483}{{\ttfamily 1805.09483}}].

\bibitem{Ballesteros:2018wlw}
G.~Ballesteros, J.~Beltrán~Jiménez and M.~Pieroni, \emph{Black hole formation
  from a general quadratic action for inflationary primordial fluctuations},
  \href{https://doi.org/10.1088/1475-7516/2019/06/016}{\emph{Journal of
  Cosmology and Astroparticle Physics} {\bfseries 2019}{\bfseries (06)} (2019)
  016} [\href{https://arxiv.org/abs/1811.03065}{{\ttfamily 1811.03065}}].

\bibitem{Georg:2019jld}
J.~Georg, B.~Melcher and S.~Watson, \emph{{Primordial black holes and
  Co-Decaying dark matter}},
  \href{https://doi.org/10.1088/1475-7516/2019/11/014}{\emph{Journal of
  Cosmology and Astroparticle Physics} {\bfseries 2019}{\bfseries (11)} (2019)
  014} [\href{https://arxiv.org/abs/1902.04082}{{\ttfamily 1902.04082}}].

\bibitem{Pi:2019ihn}
S.~Pi, M.~Sasaki and Y.-l. Zhang, \emph{Primordial tensor perturbation in
  double inflationary scenario with a break},
  \href{https://doi.org/10.1088/1475-7516/2019/06/049}{\emph{Journal of
  Cosmology and Astroparticle Physics} {\bfseries 2019}{\bfseries (06)} (2019)
  049} [\href{https://arxiv.org/abs/1904.06304}{{\ttfamily 1904.06304}}].

\bibitem{Germani:2018jgr}
C.~Germani and I.~Musco, \emph{{Abundance of Primordial Black Holes Depends on
  the Shape of the Inflationary Power Spectrum}},
  \href{https://doi.org/10.1103/PhysRevLett.122.141302}{\emph{Physical Review
  Letters} {\bfseries 122}{\bfseries (14)} (2019) 141302}
  [\href{https://arxiv.org/abs/1805.04087}{{\ttfamily 1805.04087}}].

\bibitem{Carr:2018poi}
B.~Carr and F.~Kühnel, \emph{Primordial black holes with multimodal mass
  spectra}, \href{https://doi.org/10.1103/PhysRevD.99.103535}{\emph{Physical
  Review D} {\bfseries 99}{\bfseries (10)} (2019) 103535}
  [\href{https://arxiv.org/abs/1811.06532}{{\ttfamily 1811.06532}}].

\bibitem{Kamenshchik:2018sig}
A.~Y. Kamenshchik, A.~Tronconi, T.~Vardanyan and G.~Venturi,
  \emph{Non-canonical inflation and primordial black holes production},
  \href{https://doi.org/10.1016/j.physletb.2019.02.036}{\emph{Physics Letters
  B} {\bfseries 791} (2019) 201}
  [\href{https://arxiv.org/abs/1812.02547}{{\ttfamily 1812.02547}}].

\bibitem{Hawking:1987bn}
S.~W. Hawking, \emph{Black holes from cosmic strings},
  \href{https://doi.org/10.1016/0370-2693(89)90206-2}{\emph{Physics Letters B}
  {\bfseries 231}{\bfseries (3)} (1989) 237}.

\bibitem{Garriga:1993gj}
J.~Garriga and M.~Sakellariadou, \emph{Effects of friction on cosmic strings},
  \href{https://doi.org/10.1103/PhysRevD.48.2502}{\emph{Physical Review D}
  {\bfseries 48}{\bfseries (6)} (1993) 2502}
  [\href{https://arxiv.org/abs/hep-th/9303024}{{\ttfamily hep-th/9303024}}].

\bibitem{Caldwell:1995fu}
R.~R. Caldwell and P.~Casper, \emph{Formation of black holes from collapsed
  cosmic string loops},
  \href{https://doi.org/10.1103/PhysRevD.53.3002}{\emph{Physical Review D}
  {\bfseries 53}{\bfseries (6)} (1996) 3002}
  [\href{https://arxiv.org/abs/gr-qc/9509012}{{\ttfamily gr-qc/9509012}}].

\bibitem{Jedamzik:1996mr}
K.~Jedamzik, \emph{{Primordial black hole formation during the QCD epoch}},
  \href{https://doi.org/10.1103/PhysRevD.55.R5871}{\emph{Physical Review D}
  {\bfseries 55}{\bfseries (10)} (1997) 5871}
  [\href{https://arxiv.org/abs/astro-ph/9605152}{{\ttfamily
  astro-ph/9605152}}].

\bibitem{Byrnes:2018clq}
C.~T. Byrnes, M.~Hindmarsh, S.~Young and M.~R.~S. Hawkins, \emph{{Primordial
  black holes with an accurate QCD equation of state}},
  \href{https://doi.org/10.1088/1475-7516/2018/08/041}{\emph{Journal of
  Cosmology and Astroparticle Physics} {\bfseries 2018}{\bfseries (08)} (2018)
  041} [\href{https://arxiv.org/abs/1801.06138}{{\ttfamily 1801.06138}}].

\bibitem{Chakraborty:2022mwu}
A.~Chakraborty, P.~K. Chanda, K.~L. Pandey and S.~Das, \emph{{Formation and
  Abundance of Late-forming Primordial Black Holes as Dark Matter}},
  \href{https://doi.org/10.3847/1538-4357/ac6ddd}{\emph{The Astrophysical
  Journal} {\bfseries 932}{\bfseries (2)} (2022) 119}
  [\href{https://arxiv.org/abs/2204.09628}{{\ttfamily 2204.09628}}].

\bibitem{Shandera:2018xkn}
S.~Shandera, D.~Jeong and H.~S. Grasshorn~Gebhardt, \emph{{Gravitational Waves
  from Binary Mergers of Subsolar Mass Dark Black Holes}},
  \href{https://doi.org/10.1103/PhysRevLett.120.241102}{\emph{Physical Review
  Letters} {\bfseries 120}{\bfseries (24)} (2018) 241102}
  [\href{https://arxiv.org/abs/1802.08206}{{\ttfamily 1802.08206}}].

\bibitem{Chen:2016kjx}
J.-W. Chen, J.~Liu, H.-L. Xu and Y.-F. Cai, \emph{Tracing primordial black
  holes in nonsingular bouncing cosmology},
  \href{https://doi.org/10.1016/j.physletb.2017.03.036}{\emph{Physics Letters
  B} {\bfseries 769} (2017) 561}
  [\href{https://arxiv.org/abs/1609.02571}{{\ttfamily 1609.02571}}].

\bibitem{Quintin:2016qro}
J.~Quintin and R.~H. Brandenberger, \emph{Black hole formation in a contracting
  universe}, \href{https://doi.org/10.1088/1475-7516/2016/11/029}{\emph{Journal
  of Cosmology and Astroparticle Physics} {\bfseries 2016}{\bfseries (11)}
  (2016) 029} [\href{https://arxiv.org/abs/1609.02556}{{\ttfamily
  1609.02556}}].

\bibitem{Lyth:2001nq}
D.~H. Lyth and D.~Wands, \emph{Generating the curvature perturbation without an
  inflaton}, \href{https://doi.org/10.1016/S0370-2693(01)01366-1}{\emph{Physics
  Letters B} {\bfseries 524}{\bfseries (1--2)} (2002) 5}
  [\href{https://arxiv.org/abs/hep-ph/0110002}{{\ttfamily hep-ph/0110002}}].

\bibitem{Moroi:2001ct}
T.~Moroi and T.~Takahashi, \emph{Effects of cosmological moduli fields on
  cosmic microwave background},
  \href{https://doi.org/10.1016/S0370-2693(01)01295-3}{\emph{Physics Letters B}
  {\bfseries 522}{\bfseries (3--4)} (2001) 215}
  [\href{https://arxiv.org/abs/hep-ph/0110096}{{\ttfamily hep-ph/0110096}}].

\bibitem{Enqvist:2001zp}
K.~Enqvist and M.~S. Sloth, \emph{{Adiabatic CMB perturbations in pre-Big Bang
  string cosmology}},
  \href{https://doi.org/10.1016/S0550-3213(02)00043-3}{\emph{Nuclear Physics B}
  {\bfseries 626}{\bfseries (1--2)} (2002) 395}
  [\href{https://arxiv.org/abs/hep-ph/0109214}{{\ttfamily hep-ph/0109214}}].

\bibitem{Linde:1996gt}
A.~Linde and V.~Mukhanov, \emph{{Non-Gaussian isocurvature perturbations from
  inflation}}, \href{https://doi.org/10.1103/PhysRevD.56.R535}{\emph{Physical
  Review D} {\bfseries 56}{\bfseries (2)} (1997) 535}
  [\href{https://arxiv.org/abs/astro-ph/9610219}{{\ttfamily
  astro-ph/9610219}}].

\bibitem{Mollerach:1989hu}
S.~Mollerach, \emph{Isocurvature baryon perturbations and inflation},
  \href{https://doi.org/10.1103/PhysRevD.42.313}{\emph{Physical Review D}
  {\bfseries 42}{\bfseries (2)} (1990) 313}.

\bibitem{Klimai:2012sf}
P.~A. Klimai and E.~V. Bugaev, \emph{{Primordial black hole formation from
  non-Gaussian curvature perturbations}},  in \emph{Proc. of 17th Int. Seminar
  QUARKS-2012}, vol.~2, p.~163, 2013,
  \href{https://arxiv.org/abs/1210.3262}{{\ttfamily 1210.3262}}.

\bibitem{Young:2013oia}
S.~Young and C.~T. Byrnes, \emph{{Primordial black holes in non-Gaussian
  regimes}}, \href{https://doi.org/10.1088/1475-7516/2013/08/052}{\emph{Journal
  of Cosmology and Astroparticle Physics} {\bfseries 2013}{\bfseries (08)}
  (2013) 052} [\href{https://arxiv.org/abs/1307.4995}{{\ttfamily 1307.4995}}].

\bibitem{Kawasaki:2012wr}
M.~Kawasaki, N.~Kitajima and T.~T. Yanagida, \emph{Primordial black hole
  formation from an axionlike curvaton model},
  \href{https://doi.org/10.1103/PhysRevD.87.063519}{\emph{Physical Review D}
  {\bfseries 87}{\bfseries (6)} (2013) 063519}
  [\href{https://arxiv.org/abs/1207.2550}{{\ttfamily 1207.2550}}].

\bibitem{Pi:2021dft}
S.~Pi and M.~Sasaki, \emph{{Primordial Black Hole Formation in Non-Minimal
  Curvaton Scenario}},  (2021)
  [\href{https://arxiv.org/abs/2112.12680}{{\ttfamily 2112.12680}}].

\bibitem{Meng:2022low}
D.-S. Meng, C.~Yuan and Q.-G. Huang, \emph{Primordial black holes generated by
  the non-minimal spectator field},
  \href{https://doi.org/10.1007/s11433-022-2095-5}{\emph{Science China Physics,
  Mechanics \& Astronomy} {\bfseries 66}{\bfseries (8)} (2023) 280411}
  [\href{https://arxiv.org/abs/2212.03577}{{\ttfamily 2212.03577}}].

\bibitem{Carr:2016drx}
B.~Carr, F.~Kühnel and M.~Sandstad, \emph{Primordial black holes as dark
  matter}, \href{https://doi.org/10.1103/PhysRevD.94.083504}{\emph{Physical
  Review D} {\bfseries 94}{\bfseries (8)} (2016) 083504}
  [\href{https://arxiv.org/abs/1607.06077}{{\ttfamily 1607.06077}}].

\bibitem{Garcia-Bellido:2016dkw}
J.~García-Bellido, M.~Peloso and C.~Unal, \emph{Gravitational waves at
  interferometer scales and primordial black holes in axion inflation},
  \href{https://doi.org/10.1088/1475-7516/2016/12/031}{\emph{Journal of
  Cosmology and Astroparticle Physics} {\bfseries 2016}{\bfseries (12)} (2016)
  031} [\href{https://arxiv.org/abs/1610.03763}{{\ttfamily 1610.03763}}].

\bibitem{Carr:2017edp}
B.~Carr, T.~Tenkanen and V.~Vaskonen, \emph{Primordial black holes from
  inflaton and spectator field perturbations in a matter-dominated era},
  \href{https://doi.org/10.1103/PhysRevD.96.063507}{\emph{Physical Review D}
  {\bfseries 96}{\bfseries (6)} (2017) 063507}
  [\href{https://arxiv.org/abs/1706.03746}{{\ttfamily 1706.03746}}].

\bibitem{Ezquiaga:2017fvi}
J.~M. Ezquiaga, J.~García-Bellido and E.~Ruiz~Morales, \emph{{Primordial black
  hole production in Critical Higgs Inflation}},
  \href{https://doi.org/10.1016/j.physletb.2017.11.039}{\emph{Physics Letters
  B} {\bfseries 776} (2018) 345}
  [\href{https://arxiv.org/abs/1705.04861}{{\ttfamily 1705.04861}}].

\bibitem{Espinosa:2017sgp}
J.~R. Espinosa, D.~Racco and A.~Riotto, \emph{{Cosmological Signature of the
  Standard Model Higgs Vacuum Instability: Primordial Black Holes as Dark
  Matter}},
  \href{https://doi.org/10.1103/PhysRevLett.120.121301}{\emph{Physical Review
  Letters} {\bfseries 120}{\bfseries (12)} (2018) 121301}
  [\href{https://arxiv.org/abs/1710.11196}{{\ttfamily 1710.11196}}].

\bibitem{Cable:2023lca}
A.~Cable and A.~Wilkins, \emph{{Spectators no more! How even unimportant fields
  can ruin your Primordial Black Hole model}},  (2023)
  [\href{https://arxiv.org/abs/2306.09232}{{\ttfamily 2306.09232}}].

\bibitem{Bullock:1996_Non-Gaussian}
J.~S. Bullock and J.~R. Primack, \emph{{Non-Gaussian fluctuations and
  primordial black holes from inflation}},
  \href{https://doi.org/10.1103/PhysRevD.55.7423}{\emph{Physical Review D}
  {\bfseries 55}{\bfseries (12)} (1997) 7423}
  [\href{https://arxiv.org/abs/astro-ph/9611106}{{\ttfamily
  astro-ph/9611106}}].

\bibitem{Young:2015_Influence}
S.~Young, D.~Regan and C.~T. Byrnes, \emph{Influence of large local and
  non-local bispectra on primordial black hole abundance},
  \href{https://doi.org/10.1088/1475-7516/2016/02/029}{\emph{Journal of
  Cosmology and Astroparticle Physics} {\bfseries 2016}{\bfseries (02)} (2016)
  029} [\href{https://arxiv.org/abs/1512.07224}{{\ttfamily 1512.07224}}].

\bibitem{Yoo:2019_Abundance}
C.-M. Yoo, J.-O. Gong and S.~Yokoyama, \emph{{Abundance of primordial black
  holes with local non-Gaussianity in peak theory}},
  \href{https://doi.org/10.1088/1475-7516/2019/09/033}{\emph{Journal of
  Cosmology and Astroparticle Physics} {\bfseries 2019}{\bfseries (09)} (2019)
  033} [\href{https://arxiv.org/abs/1906.06790}{{\ttfamily 1906.06790}}].

\bibitem{Taoso:2021_Non-gaussianities}
M.~Taoso and A.~Urbano, \emph{Non-gaussianities for primordial black hole
  formation},
  \href{https://doi.org/10.1088/1475-7516/2021/08/016}{\emph{Journal of
  Cosmology and Astroparticle Physics} {\bfseries 2021}{\bfseries (08)} (2021)
  016} [\href{https://arxiv.org/abs/2102.03610}{{\ttfamily 2102.03610}}].

\bibitem{Young:2022_non-G}
S.~Young, \emph{{Peaks and primordial black holes: the effect of
  non-Gaussianity}},
  \href{https://doi.org/10.1088/1475-7516/2022/05/037}{\emph{Journal of
  Cosmology and Astroparticle Physics} {\bfseries 2022}{\bfseries (05)} (2022)
  037} [\href{https://arxiv.org/abs/2201.13345}{{\ttfamily 2201.13345}}].

\bibitem{Gow:2022jfb}
A.~D. Gow \textit{et~al.}, \emph{{Non-perturbative non-Gaussianity and
  primordial black holes}},
  \href{https://doi.org/10.1209/0295-5075/acd417}{\emph{Europhysics Letters}
  {\bfseries 142}{\bfseries (4)} (2023) 49001}
  [\href{https://arxiv.org/abs/2211.08348}{{\ttfamily 2211.08348}}].

\bibitem{Ferrante:2022mui}
G.~Ferrante, G.~Franciolini, A.~J. Iovino and A.~Urbano, \emph{{Primordial
  non-Gaussianity up to all orders: Theoretical aspects and implications for
  primordial black hole models}},
  \href{https://doi.org/10.1103/PhysRevD.107.043520}{\emph{Physical Review D}
  {\bfseries 107}{\bfseries (4)} (2023) 043520}
  [\href{https://arxiv.org/abs/2211.01728}{{\ttfamily 2211.01728}}].

\bibitem{Ferrante:2023bgz}
G.~Ferrante, G.~Franciolini, A.~J. Iovino and A.~Urbano, \emph{Primordial black
  holes in the curvaton model: possible connections to pulsar timing arrays and
  dark matter},  (2023) [\href{https://arxiv.org/abs/2305.13382}{{\ttfamily
  2305.13382}}].

\bibitem{Starobinsky:1994bd}
A.~A. Starobinsky and J.~Yokoyama, \emph{{Equilibrium state of a
  self-interacting scalar field in the de Sitter background}},
  \href{https://doi.org/10.1103/PhysRevD.50.6357}{\emph{Physical Review D}
  {\bfseries 50}{\bfseries (10)} (1994) 6357}
  [\href{https://arxiv.org/abs/astro-ph/9407016}{{\ttfamily
  astro-ph/9407016}}].

\bibitem{Enqvist:2012xn}
K.~Enqvist, R.~N. Lerner, O.~Taanila and A.~Tranberg, \emph{{Spectator field
  dynamics in de Sitter and curvaton initial conditions}},
  \href{https://doi.org/10.1088/1475-7516/2012/10/052}{\emph{Journal of
  Cosmology and Astroparticle Physics} {\bfseries 2012}{\bfseries (10)} (2012)
  052} [\href{https://arxiv.org/abs/1205.5446}{{\ttfamily 1205.5446}}].

\bibitem{Hardwick:2017fjo}
R.~J. Hardwick \textit{et~al.}, \emph{The stochastic spectator},
  \href{https://doi.org/10.1088/1475-7516/2017/10/018}{\emph{Journal of
  Cosmology and Astroparticle Physics} {\bfseries 2017}{\bfseries (10)} (2017)
  018} [\href{https://arxiv.org/abs/1701.06473}{{\ttfamily 1701.06473}}].

\bibitem{Planck:2019kim}
{\scshape Planck} collaboration, \emph{{Planck 2018 results. IX. Constraints on
  primordial non-Gaussianity}},
  \href{https://doi.org/10.1051/0004-6361/201935891}{\emph{Astronomy \&
  Astrophysics} {\bfseries 641} (2020) A9}
  [\href{https://arxiv.org/abs/1905.05697}{{\ttfamily 1905.05697}}].

\bibitem{Starobinsky:1985ibc}
A.~A. Starobinsky, \emph{{Multicomponent de Sitter (inflationary) stages and
  the generation of perturbations}},
  \href{http://www.jetpletters.ru/ps/1419/article_21563.shtml}{\emph{Journal of
  Experimental and Theoretical Physics Letters} {\bfseries 42}{\bfseries (3)}
  (1985) 124}.

\bibitem{Salopek:1990jq}
D.~S. Salopek and J.~R. Bond, \emph{Nonlinear evolution of long-wavelength
  metric fluctuations in inflationary models},
  \href{https://doi.org/10.1103/PhysRevD.42.3936}{\emph{Physical Review D}
  {\bfseries 42}{\bfseries (12)} (1990) 3936}.

\bibitem{Sasaki:1995aw}
M.~Sasaki and E.~D. Stewart, \emph{{A General Analytic Formula for the Spectral
  Index of the Density Perturbations Produced during Inflation}},
  \href{https://doi.org/10.1143/PTP.95.71}{\emph{Progress of Theoretical
  Physics} {\bfseries 95}{\bfseries (1)} (1996) 71}
  [\href{https://arxiv.org/abs/astro-ph/9507001}{{\ttfamily
  astro-ph/9507001}}].

\bibitem{Wands:2000dp}
D.~Wands, K.~A. Malik, D.~H. Lyth and A.~R. Liddle, \emph{New approach to the
  evolution of cosmological perturbations on large scales},
  \href{https://doi.org/10.1103/PhysRevD.62.043527}{\emph{Physical Review D}
  {\bfseries 62}{\bfseries (4)} (2000) 043527}
  [\href{https://arxiv.org/abs/astro-ph/0003278}{{\ttfamily
  astro-ph/0003278}}].

\bibitem{Musco:2008hv}
I.~Musco, J.~C. Miller and A.~G. Polnarev, \emph{Primordial black hole
  formation in the radiative era: investigation of the critical nature of the
  collapse},
  \href{https://doi.org/10.1088/0264-9381/26/23/235001}{\emph{Classical and
  Quantum Gravity} {\bfseries 26}{\bfseries (3)} (2009) 235001}
  [\href{https://arxiv.org/abs/0811.1452}{{\ttfamily 0811.1452}}].

\bibitem{Musco:2020jjb}
I.~Musco, V.~De~Luca, G.~Franciolini and A.~Riotto, \emph{{Threshold for
  primordial black holes. II. A simple analytic prescription}},
  \href{https://doi.org/10.1103/PhysRevD.103.063538}{\emph{Physical Review D}
  {\bfseries 103}{\bfseries (6)} (2021) 063538}
  [\href{https://arxiv.org/abs/2011.03014}{{\ttfamily 2011.03014}}].

\bibitem{Ebadi:2023xhq}
R.~Ebadi \textit{et~al.}, \emph{{Gravitational Waves from Stochastic Scalar
  Fluctuations}},  (2023) [\href{https://arxiv.org/abs/2307.01248}{{\ttfamily
  2307.01248}}].

\bibitem{Bunch:1978yq}
T.~S. Bunch and P.~C.~W. Davies, \emph{{Quantum field theory in de Sitter
  space: renormalization by point splitting}},
  \href{https://doi.org/10.1098/rspa.1978.0060}{\emph{Proceedings of the Royal
  Society A} {\bfseries 360}{\bfseries (1700)} (1978) 117}.

\bibitem{Sasaki:2006kq}
M.~Sasaki, J.~Väliviita and D.~Wands, \emph{{Non-Gaussianity of the primordial
  perturbation in the curvaton model}},
  \href{https://doi.org/10.1103/PhysRevD.74.103003}{\emph{Physical Review D}
  {\bfseries 74}{\bfseries (10)} (2006) 103003}
  [\href{https://arxiv.org/abs/astro-ph/0607627}{{\ttfamily
  astro-ph/0607627}}].

\bibitem{Planck:2018vyg}
{\scshape Planck} collaboration, \emph{{Planck 2018 results. VI. Cosmological
  parameters}},
  \href{https://doi.org/10.1051/0004-6361/201833910}{\emph{Astronomy \&
  Astrophysics} {\bfseries 641} (2020) A6}
  [\href{https://arxiv.org/abs/1807.06209}{{\ttfamily 1807.06209}}].

\bibitem{Tristram:2020wbi}
M.~Tristram \textit{et~al.}, \emph{Planck constraints on the tensor-to-scalar
  ratio}, \href{https://doi.org/10.1051/0004-6361/202039585}{\emph{Astronomy \&
  Astrophysics} {\bfseries 647} (2021) A128}
  [\href{https://arxiv.org/abs/2010.01139}{{\ttfamily 2010.01139}}].

\bibitem{Shibata:1999zs}
M.~Shibata and M.~Sasaki, \emph{{Black hole formation in the Friedmann
  universe: Formulation and computation in numerical relativity}},
  \href{https://doi.org/10.1103/PhysRevD.60.084002}{\emph{Physical Review D}
  {\bfseries 60}{\bfseries (8)} (1999) 084002}
  [\href{https://arxiv.org/abs/gr-qc/9905064}{{\ttfamily gr-qc/9905064}}].

\bibitem{Harada:2015yda}
T.~Harada, C.-M. Yoo, T.~Nakama and Y.~Koga, \emph{Cosmological long-wavelength
  solutions and primordial black hole formation},
  \href{https://doi.org/10.1103/PhysRevD.91.084057}{\emph{Physical Review D}
  {\bfseries 91}{\bfseries (8)} (2015) 084057}
  [\href{https://arxiv.org/abs/1503.03934}{{\ttfamily 1503.03934}}].

\bibitem{Yoo:2018kvb}
C.-M. Yoo, T.~Harada, J.~Garriga and K.~Kohri, \emph{{Primordial black hole
  abundance from random Gaussian curvature perturbations and a local density
  threshold}}, \href{https://doi.org/10.1093/ptep/pty120}{\emph{Progress of
  Theoretical and Experimental Physics} {\bfseries 2018}{\bfseries (12)} (2018)
  123E01} [\href{https://arxiv.org/abs/1805.03946}{{\ttfamily 1805.03946}}].

\bibitem{Musco:2018rwt}
I.~Musco, \emph{{Threshold for primordial black holes: Dependence on the shape
  of the cosmological perturbations}},
  \href{https://doi.org/10.1103/PhysRevD.100.123524}{\emph{Physical Review D}
  {\bfseries 100}{\bfseries (12)} (2019) 123524}
  [\href{https://arxiv.org/abs/1809.02127}{{\ttfamily 1809.02127}}].

\bibitem{Young:2019yug}
S.~Young, I.~Musco and C.~T. Byrnes, \emph{Primordial black hole formation and
  abundance: contribution from the non-linear relation between the density and
  curvature perturbation},
  \href{https://doi.org/10.1088/1475-7516/2019/11/012}{\emph{Journal of
  Cosmology and Astroparticle Physics} {\bfseries 2019}{\bfseries (11)} (2019)
  012} [\href{https://arxiv.org/abs/1904.00984}{{\ttfamily 1904.00984}}].

\bibitem{Choptuik:1992jv}
M.~W. Choptuik, \emph{Universality and scaling in gravitational collapse of a
  massless scalar field},
  \href{https://doi.org/10.1103/PhysRevLett.70.9}{\emph{Physical Review
  Letters} {\bfseries 70}{\bfseries (1)} (1993) 9}.

\bibitem{Evans:1994pj}
C.~R. Evans and J.~S. Coleman, \emph{Critical phenomena and self-similarity in
  the gravitational collapse of radiation fluid},
  \href{https://doi.org/10.1103/PhysRevLett.72.1782}{\emph{Physical Review
  Letters} {\bfseries 72}{\bfseries (12)} (1994) 1782}
  [\href{https://arxiv.org/abs/gr-qc/9402041}{{\ttfamily gr-qc/9402041}}].

\bibitem{Niemeyer:1997mt}
J.~C. Niemeyer and K.~Jedamzik, \emph{{Near-Critical Gravitational Collapse and
  the Initial Mass Function of Primordial Black Holes}},
  \href{https://doi.org/10.1103/PhysRevLett.80.5481}{\emph{Physical Review
  Letters} {\bfseries 80}{\bfseries (25)} (1998) 5481}
  [\href{https://arxiv.org/abs/astro-ph/9709072}{{\ttfamily
  astro-ph/9709072}}].

\bibitem{Niemeyer:1999ak}
J.~C. Niemeyer and K.~Jedamzik, \emph{Dynamics of primordial black hole
  formation}, \href{https://doi.org/10.1103/PhysRevD.59.124013}{\emph{Physical
  Review D} {\bfseries 59}{\bfseries (12)} (1999) 124013}
  [\href{https://arxiv.org/abs/astro-ph/9901292}{{\ttfamily
  astro-ph/9901292}}].

\bibitem{Musco:2004ak}
I.~Musco, J.~C. Miller and L.~Rezzolla, \emph{Computations of primordial
  black-hole formation},
  \href{https://doi.org/10.1088/0264-9381/22/7/013}{\emph{Classical and Quantum
  Gravity} {\bfseries 22}{\bfseries (7)} (2005) 1405}
  [\href{https://arxiv.org/abs/gr-qc/0412063}{{\ttfamily gr-qc/0412063}}].

\bibitem{Musco:2012au}
I.~Musco and J.~C. Miller, \emph{Primordial black hole formation in the early
  universe: critical behaviour and self-similarity},
  \href{https://doi.org/10.1088/0264-9381/30/14/145009}{\emph{Classical and
  Quantum Gravity} {\bfseries 30}{\bfseries (14)} (2013) 145009}
  [\href{https://arxiv.org/abs/1201.2379}{{\ttfamily 1201.2379}}].

\bibitem{Young:2019osy}
S.~Young, \emph{{The primordial black hole formation criterion re-examined:
  Parametrisation, timing and the choice of window function}},
  \href{https://doi.org/10.1142/S0218271820300025}{\emph{International Journal
  of Modern Physics D} {\bfseries 29}{\bfseries (02)} (2020) 2030002}
  [\href{https://arxiv.org/abs/1905.01230}{{\ttfamily 1905.01230}}].

\bibitem{Escriva:2020tak}
A.~Escrivà, C.~Germani and R.~K. Sheth, \emph{Analytical thresholds for black
  hole formation in general cosmological backgrounds},
  \href{https://doi.org/10.1088/1475-7516/2021/01/030}{\emph{Journal of
  Cosmology and Astroparticle Physics} {\bfseries 2021}{\bfseries (01)} (2021)
  030} [\href{https://arxiv.org/abs/2007.05564}{{\ttfamily 2007.05564}}].

\bibitem{Kopp:2010sh}
M.~Kopp, S.~Hofmann and J.~Weller, \emph{Separate universes do not constrain
  primordial black hole formation},
  \href{https://doi.org/10.1103/PhysRevD.83.124025}{\emph{Physical Review D}
  {\bfseries 83}{\bfseries (12)} (2011) 124025}
  [\href{https://arxiv.org/abs/1012.4369}{{\ttfamily 1012.4369}}].

\bibitem{Starobinsky:1986fx}
A.~A. Starobinsky, \emph{Stochastic de sitter (inflationary) stage in the early
  universe},  in \emph{Lecture Notes in Physics (Field Theory, Quantum Gravity
  and Strings)}, vol.~246, p.~107, 1988,
  \href{https://doi.org/10.1007/3-540-16452-9\_6}{10.1007/3-540-16452-9\_6}.

\bibitem{Markkanen:2019kpv}
T.~Markkanen, A.~Rajantie, S.~Stopyra and T.~Tenkanen, \emph{{Scalar
  correlation functions in de Sitter space from the stochastic spectral
  expansion}},
  \href{https://doi.org/10.1088/1475-7516/2019/08/001}{\emph{Journal of
  Cosmology and Astroparticle Physics} {\bfseries 2019}{\bfseries (08)} (2019)
  001} [\href{https://arxiv.org/abs/1904.11917}{{\ttfamily 1904.11917}}].

\bibitem{Markkanen:2020bfc}
T.~Markkanen and A.~Rajantie, \emph{{Scalar correlation functions for a
  double-well potential in de Sitter space}},
  \href{https://doi.org/10.1088/1475-7516/2020/03/049}{\emph{Journal of
  Cosmology and Astroparticle Physics} {\bfseries 2020}{\bfseries (03)} (2020)
  049} [\href{https://arxiv.org/abs/2001.04494}{{\ttfamily 2001.04494}}].

\bibitem{Langlois:2004nn}
D.~Langlois and F.~Vernizzi, \emph{Mixed inflaton and curvaton perturbations},
  \href{https://doi.org/10.1103/PhysRevD.70.063522}{\emph{Physical Review D}
  {\bfseries 70}{\bfseries (6)} (2004) 063522}
  [\href{https://arxiv.org/abs/astro-ph/0403258}{{\ttfamily
  astro-ph/0403258}}].

\end{thebibliography}\endgroup
\bibliographystyle{JHEP-edit}

\end{document}